\documentstyle[12pt]{article}
\begin{document}

\def\ds{\displaystyle}
\def\beq{\begin{equation}}
\def\eeq{\end{equation}}
\def\bea{\begin{eqnarray}}
\def\eea{\end{eqnarray}}
\def\beeq{\begin{eqnarray}}
\def\eeeq{\end{eqnarray}}
\def\ve{\vert}
\def\vel{\left|}
\def\ver{\right|}
\def\nnb{\nonumber}
\def\ga{\left(}
\def\dr{\right)}
\def\aga{\left\{}
\def\adr{\right\}}
\def\lla{\left<}
\def\rra{\right>}
\def\rar{\rightarrow}
\def\nnb{\nonumber}
\def\la{\langle}
\def\ra{\rangle}
\def\ba{\begin{array}}
\def\ea{\end{array}}
\def\tr{\mbox{Tr}}
\def\ssp{{\Sigma^{*+}}}
\def\sso{{\Sigma^{*0}}}
\def\ssm{{\Sigma^{*-}}}
\def\xis0{{\Xi^{*0}}}
\def\xism{{\Xi^{*-}}}
\def\qs{\la \bar s s \ra}
\def\qu{\la \bar u u \ra}
\def\qd{\la \bar d d \ra}
\def\qq{\la \bar q q \ra}
\def\gGgG{\la g^2 G^2 \ra}
\def\q{\gamma_5 \not\!q}
\def\x{\gamma_5 \not\!x}
\def\g5{\gamma_5}
\def\sb{S_Q^{cf}}
\def\sd{S_d^{be}}
\def\su{S_u^{ad}}
\def\ss{S_s^{??}}
\def\sbp{{S}_Q^{'cf}}
\def\sdp{{S}_d^{'be}}
\def\sup{{S}_u^{'ad}}
\def\ssp{{S}_s^{'??}}
\def\sig{\sigma_{\mu \nu} \gamma_5 p^\mu q^\nu}
\def\fo{f_0(\frac{s_0}{M^2})}
\def\ffi{f_1(\frac{s_0}{M^2})}
\def\fii{f_2(\frac{s_0}{M^2})}
\def\O{{\cal O}}
\def\sl{{\Sigma^0 \Lambda}}
\def\es{\!\!\! &=& \!\!\!}
\def\ar{&+& \!\!\!}
\def\ek{&-& \!\!\!}
\def\cp{&\times& \!\!\!}
\def\se{\!\!\! &\simeq& \!\!\!}


\renewcommand{\textfraction}{0.2}    
\renewcommand{\topfraction}{0.8}   

\renewcommand{\bottomfraction}{0.4}   
\renewcommand{\floatpagefraction}{0.8}
\newcommand\mysection{\setcounter{equation}{0}\section}

\def\baeq{\begin{appeq}}     \def\eaeq{\end{appeq}}  
\def\baeeq{\begin{appeeq}}   \def\eaeeq{\end{appeeq}}
\newenvironment{appeq}{\beq}{\eeq}   
\newenvironment{appeeq}{\beeq}{\eeeq}
\def\bAPP#1#2{
 \markright{APPENDIX #1}
 \addcontentsline{toc}{section}{Appendix #1: #2}
 \medskip
 \medskip
 \begin{center}      {\bf\LARGE Appendix #1 :}{\quad\Large\bf #2}
\end{center}
 \renewcommand{\thesection}{#1.\arabic{section}}
\setcounter{equation}{0}
        \renewcommand{\thehran}{#1.\arabic{hran}}
\renewenvironment{appeq}
  {  \renewcommand{\theequation}{#1.\arabic{equation}}
     \beq
  }{\eeq}
\renewenvironment{appeeq}
  {  \renewcommand{\theequation}{#1.\arabic{equation}}
     \beeq
  }{\eeeq}
\nopagebreak \noindent}

\def\eAPP{\renewcommand{\thehran}{\thesection.\arabic{hran}}}

\renewcommand{\theequation}{\arabic{equation}}
\newcounter{hran}
\renewcommand{\thehran}{\thesection.\arabic{hran}}

\def\bmini{\setcounter{hran}{\value{equation}}
\refstepcounter{hran}\setcounter{equation}{0}
\renewcommand{\theequation}{\thehran\alph{equation}}\begin{eqnarray}}
\def\bminiG#1{\setcounter{hran}{\value{equation}}
\refstepcounter{hran}\setcounter{equation}{-1}
\renewcommand{\theequation}{\thehran\alph{equation}}
\refstepcounter{equation}\label{#1}\begin{eqnarray}}


\newskip\humongous \humongous=0pt plus 1000pt minus 1000pt
\def\caja{\mathsurround=0pt}


\title{
         {\Large
                 {\bf
Exclusive $\Lambda_b \rar \Lambda \ell^+ \ell^-$ decay beyond standard model
                 }
         }
      }

\author{\vspace{1cm}\\
{\small T. M. Aliev$^a$ \thanks
{e-mail: taliev@metu.edu.tr}\,\,,
A. \"{O}zpineci$^b$ \thanks
{e-mail: ozpineci@ictp.trieste.it}\,\,,
M. Savc{\i}$^a$ \thanks
{e-mail: savci@metu.edu.tr}} \\
{\small a Physics Department, Middle East Technical University, 
06531 Ankara, Turkey}\\
{\small b  The Abdus Salam International Center for Theoretical Physics,
I-34100, Trieste, Italy} }
\date{}

\begin{titlepage}
\maketitle
\thispagestyle{empty}

\begin{abstract}
Using the most general, model independent form of the effective Hamiltonian,
the exclusive, rare baryonic $\Lambda_b \rar \Lambda \ell^+
\ell^-~(\ell=\mu,\tau)$ decay is analyzed. We study sensitivity of the
branching ratio and lepton forward--backward asymmetry to the new Wilson
coefficients. Is is shown that these physical quantities are quite sensitive
to the new Wilson coefficients. Determination of the position of zero
value of the forward--backward asymmetry can serve as a useful tool for
establishing new physics beyond the standard model, as well as fixing the
sign of the new Wilson coefficients.   
\end{abstract}

~~~PACS numbers: 12.60.--i, 13.30.--a, 14.20.Mr
\end{titlepage}

\section{Introduction}
Rare decays, induced by flavor changing neutral current (FCNC) $b \rar s(d)$
transitions, provide testing ground for the standard model (SM) at loop
level. For this reason studying these decays constitute one of the main
research directions of the two operating $B$--factories BaBar and Belle
\cite{R461}. Rare decays can give valuable information about poorly studied
aspects of the SM at present, such as
Cabibbo--Kobayashi--Maskawa matrix elements, $V_{td},~V_{ts}$, and $V_{ub}$
and leptonic decay constant.
After CLEO measurement of the radiative decay $b \rar s \gamma$ decay
\cite{R462}, main interest has been focused on the rare decays induced by the 
$b \rar s \ell^+ \ell^-$ transition, which have relatively "large" branching
ratio in the SM. These decays have been investigated extensively
in the SM and its various extensions \cite{R463}--\cite{R4618}.

The theoretical analysis of the inclusive decays is rather easy since they
are free of long distance effects, but their experimental detection is quite
difficult. For exclusive decays the situation is contrary to the inclusive
case; i.e., their experimental investigation is easy, but theoretical
analysis is difficult due to the appearance of the form factors. It should
be noted that the exclusive $B\rar K^\ast(K) \ell^+ \ell^-$ decays, which
are described by the $b \rar s \ell^+ \ell^-$ transition at inclusive level,  
have been widely studied in literature (see \cite{R4619}--\cite{R4622} and 
references therein). Another exclusive decay which is described at inclusive 
level by the $b \rar s \ell^+ \ell^-$ transition is the baryonic $\Lambda_b \rar
\Lambda \ell^+\ell^-$ decay. This decay has been studied in context of the
SM and two Higgs doublet models in \cite{R4623} and \cite{R4624},
respectively. 

Rare decays are very sensitive to the new physics beyond the SM and
therefore constitute quite a suitable tool for looking such effects. In
general, new physics effects manifest themselves in rare decays either
through new contributions to the Wilson coefficients existing in the SM or
by introducing new structures in the effective Hamiltonian which are absent
in the SM (see for example \cite{R4621}, \cite{R4625}--\cite{R4627} and the
references therein). At this point we would like to remind that, the
sensitivity of the physical observables to the new physics effects in the
"heavy pseudoscalar meson $\rar$ light pseudoscalar (vector) meson" transitions,
like $B \rar K(K^\ast) \ell^+ \ell^-$ are studied systematically in 
\cite{R4621,R4627,R4628}, using the most general form of the effective
Hamiltonian. 

The intriguing questions that follow next are what happens in the 
"heavy baryon $\rar$ light baryon" transition and which physical quantity is most
sensitive to the new physics effects. The present work is devoted to find an
answer to these questions.  

In this work we present a systematic study of the baryonic $\Lambda_b \rar
\Lambda \ell^+\ell^-$ decay. The paper is organized as follows. In section
2, using the most general model independent form of the Hamiltonian we
derive the matrix element, differential decay width and forward--backward
asymmetry, in terms of the form factors. Section 3 is devoted to the
numerical analysis and concluding remarks.  

\section{Theoretical background}

The matrix element of the $\Lambda_b \rar\Lambda \ell^+\ell^-$ decay at
quark level is described by the $b \rar s \ell^+\ell^-$ transition. The
decay amplitude for the $b \rar s \ell^+\ell^-$ transition in a general
model independent form can be written in the following way (see
\cite{R4621},\cite{R4625,R4626})
\bea
\label{e1}
\lefteqn{
{\cal M} = \frac{G \alpha}{4\sqrt{2} \pi} V_{tb}V_{ts}^\ast \Bigg\{
C_{SL} \bar s i \sigma_{\mu\nu} \frac{q^\nu}{q^2} L b \bar\ell \gamma_\mu \ell +
C_{BR} \bar s i \sigma_{\mu\nu} \frac{q^\nu}{q^2} b \bar\ell \gamma_\mu \ell +
C_{LL}^{tot} \bar s_L \gamma^\mu b_L \bar \ell_L\gamma_\mu \ell_L} \nnb\\
\ar C_{LR}^{tot} \bar s_L \gamma^\mu b_L \bar \ell_R \gamma_\mu \ell_R +
C_{RL} \bar s_R \gamma^\mu b_R \bar \ell_L \gamma_\mu \ell_L +
C_{RR} \bar s_R \gamma^\mu b_R \bar \ell_R \gamma_\mu \ell_R \nnb \\
\ar C_{LRLR} \bar s_L b_R \bar \ell_L \ell_R +
C_{RLLR} \bar s_R b_L \bar \ell_L \ell_R +
C_{LRRL} \bar s_L b_R \bar \ell_R \ell_L +
C_{RLRL} \bar s_R b_L \bar \ell_R \ell_L \nnb \\
\ar C_T \bar s \sigma^{\mu\nu} b \bar \ell \sigma_{\mu\nu} \ell +
i C_{TE} \epsilon^{\mu\nu\alpha\beta} \bar s \sigma_{\mu\nu} s
\sigma_{\alpha\beta} \ell \Bigg\}~,
\eea
where $L=(1-\gamma_5)/2$ and $R=(1+\gamma_5)/2$ are the chiral operators and
$C_X$ are the coefficients of the four--Fermi interaction. Part of these
coefficients exist in the SM. The first two of the coefficients $C_{SL}$ and 
$C_{BR}$ are the nonlocal Fermi interactions which correspond to $-2 m_s
C_7^{eff}$ and $-2 m_b C_7^{eff}$ in the SM, respectively. The following
four terms describe vector type interactions. Two of these vector
interactions containing coefficients $C_{LL}^{tot}$ and $C_{LR}^{tot}$ do
also exist in the SM in the forms $(C_9^{eff}-C_{10})$ and
$(C_9^{eff}+C_{10})$, respectively. Therefore $C_{LL}^{tot}$ and
$C_{LR}^{tot}$ represent the sum of the combinations from SM and the new
physics, in the following forms 
\bea
\label{e2}
C_{LL}^{tot} \es C_9^{eff}- C_{10} + C_{LL}~, \nnb \\
C_{LR}^{tot} \es C_9^{eff}+ C_{10} + C_{LL}~.
\eea
The terms with  $C_{LRRL},~C_{LRLR},~C_{RLRL}$ and $C_{RLLR}$ describe the
scalar type interactions. The last two terms in Eq. (\ref{e1}) correspond to
the tensor type interactions. The amplitude of the exclusive $\Lambda_b
\rar\Lambda \ell^+\ell^-$ decay can be obtained sandwiching matrix
element of the $b \rar s \ell^+ \ell^-$ decay between initial and final
state baryons. It follows from Eq. (\ref{e1}) that, in order to calculate
the amplitude of the $\Lambda_b \rar\Lambda \ell^+\ell^-$ decay the
following matrix elements are needed 
\bea
\label{e3}
&&\lla \Lambda \vel \bar s \gamma_\mu (1 \mp \gamma_5) b \ver \Lambda_b
\rra~,\nnb \\
&&\lla \Lambda \vel \bar s \sigma_{\mu\nu} (1 \mp \gamma_5) b \ver \Lambda_b
\rra~,\nnb \\
&&\lla \Lambda \vel \bar s (1 \mp \gamma_5) b \ver \Lambda_b \rra~.
\eea
Explicit forms of these matrix elements in terms of the form factors are
presented in 
Appendix--A. Using the parametrization of these matrix elements, the matrix
form of the $\Lambda_b \rar\Lambda \ell^+\ell^-$ decay can be written as 
\bea
\label{e4}
\lefteqn{
{\cal M} = \frac{G \alpha}{4 \sqrt{2}\pi} V_{tb}V_{ts}^\ast \Bigg\{
\bar \ell \gamma^\mu \ell \, \bar u_\Lambda \Big[ A_1 \gamma_\mu (1+\gamma_5) +
B_1 \gamma_\mu (1-\gamma_5) }\nnb \\
\ar i \sigma_{\mu\nu} q^\nu \big[ A_2 (1+\gamma_5) + B_2 (1-\gamma_5) \big]
+q_\mu \big[ A_3 (1+\gamma_5) + B_3 (1-\gamma_5) \big]\Big] u_{\Lambda_b}
\nnb \\
\ar \bar \ell \gamma^\mu \gamma_5 \ell \, \bar u_\Lambda \Big[
D_1 \gamma_\mu (1+\gamma_5) + E_1 \gamma_\mu (1-\gamma_5) +
i \sigma_{\mu\nu} q^\nu \big[ D_2 (1+\gamma_5) + E_2 (1-\gamma_5) \big]
\nnb \\
\ar q_\mu \big[ D_3 (1+\gamma_5) + E_3 (1-\gamma_5) \big]\Big] u_{\Lambda_b}+
\bar \ell \ell\, \bar u_\Lambda \big(N_1 + H_1 \gamma_5\big) u_{\Lambda_b}
+\bar \ell \gamma_5 \ell \, \bar u_\Lambda \big(N_2 + H_2 \gamma_5\big) 
u_{\Lambda_b}\nnb \\
\ar 4 C_T \bar \ell \sigma^{\mu\nu}\ell \, \bar u_\Lambda \Big[ f_T 
\sigma_{\mu\nu} - i f_T^V \big( q_\nu \gamma_\mu - q_\mu \gamma_\nu \big) -
i f_T^S \big( P_\mu q_\nu - P_\nu q_\mu \big) \Big] u_{\Lambda_b}\nnb \\
\ar 4 C_{TE} \epsilon^{\mu\nu\alpha\beta} \bar \ell \sigma_{\alpha\beta}
\ell \, i \bar u_\Lambda \Big[ f_T \sigma_{\mu\nu} - 
i f_T^V \big( q_\nu \gamma_\mu - q_\mu \gamma_\nu \big) -
i f_T^S \big( P_\mu q_\nu - P_\nu q_\mu \big) \Big] u_{\Lambda_b}\Bigg\}~,
\eea
where $P=p_{\Lambda_b}+ p_\Lambda$.

Explicit expressions of the functions $A_i,~B_i,~D_i,~E_i,~H_j$ and $N_j$
$(i=1,2,3$ and $j=1,2)$ are given in Appendix--A. 

Obviously, the $\Lambda_b \rar\Lambda \ell^+\ell^-$ decay introduces a lot
of form factors. However, when the heavy quark effective theory (HQET) has
been used, the heavy quark symmetry reduces the number of independent form
factors to two only $(F_1$ and $F_2)$, irrelevant with the Dirac structure of
the relevant operators \cite{R4629}, and hence we obtain that
\bea
\label{e5}
\lla \Lambda(p_\Lambda) \vel \bar s \Gamma b \ver \Lambda(p_{\Lambda_b})
\rra = \bar u_\Lambda \Big[F_1(q^2) + \not\!v F_2(q^2)\Big] \Gamma
u_{\Lambda_b}~,
\eea
where $\Gamma$ is an arbitrary Dirac structure,
$v^\mu=p_{\Lambda_b}^\mu/m_{\Lambda_b}$ is the four--velocity of
$\Lambda_b$, and $q=p_{\Lambda_b}-p_\Lambda$ is the momentum transfer.
Comparing the general form of the form factors with (\ref{e5}), one can
easily obtain the following relations among them (see also \cite{R4623})
\bea
\label{e6}
g_1 \es f_1 = f_2^T= g_2^T = F_1 + \sqrt{r} F_2~, \nnb \\
g_2 \es f_2 = g_3 = f_3 = g_T^V = f_T^V = \frac{F_2}{m_{\Lambda_b}}~,\nnb \\
g_T^S \es f_T^S = 0 ~,\nnb \\
g_1^T \es f_1^T = \frac{F_2}{m_{\Lambda_b}} q^2~,\nnb \\
g_3^T \es \frac{F_2}{m_{\Lambda_b}} \ga m_{\Lambda_b} + m_\Lambda \dr~,\nnb \\
f_3^T \es - \frac{F_2}{m_{\Lambda_b}} \ga m_{\Lambda_b} - m_\Lambda \dr~,
\eea
where $r=m_\Lambda^2/m_{\Lambda_b}^2$. 
These relations will be used in further numerical calculations.

It is a simple matter now to derive the double differential rate with
respect to the angle between lepton and the dimensionless 
invariant mass of the dilepton
\bea
\label{e7} 
\frac{d^2\Gamma}{dsdz} = \frac{G^2\alpha^2 m_{\Lambda_b}}{16384 \pi^5}
\vel V_{tb}V_{ts}^\ast \ver^2 v \sqrt{\lambda(1,r,s)} \, {\cal T}(s,z)~,
\eea     
where $s = q^2/m_{\Lambda_b}^2$ and 
\bea
\label{e8}
{\cal T}(s,z) = {\cal T}_0(s)+{\cal T}_1(s)z+{\cal T}_2(s)z^2~.
\eea
The expressions for ${\cal T}_0(s)$, ${\cal T}_1(s)$ and 
${\cal T}_2(s)$ can be found in Appendix--B. 

In Eqs. (\ref{e7})--(\ref{e8}), $z=\cos\theta$ is the angle between the
momenta of $\ell^-$ and $\Lambda_b$ in the center of mass frame of
dileptons, $\lambda(1,r,s)-1+r^2+s^2-2r-2s-2rs$ is the triangle function. 
After integrating over the angle $z$, the invariant dilepton mass distribution
takes the following form
\bea
\label{e12}
\frac{d\Gamma}{ds} = \frac{G^2\alpha^2 m_{\Lambda_b}}{8192 \pi^5}\vel
V_{tb}V_{ts}^\ast\ver^2 v \sqrt{\lambda(1,r,s)} \, \Bigg[ {\cal T}_0(s) +
\frac{1}{3}{\cal T}_2(s)\Bigg]~.
\eea
The limit for $s$ is given by
\bea
\label{e13}
\frac{4 m_\ell^2}{m_{\Lambda_b}^2} \le s \le (1-\sqrt{r})^2~.
\eea

The lepton forward--backward asymmetry ${\cal A}_{FB}$ is one of the powerful 
tools in looking for new physics beyond the SM. The determination of the 
position of the zero value of the ${\cal A}_{FB}$ is very useful for this
purpose. The new physics effects can shift the position of the zero value of
the forward--backward asymmetry. Indeed, it has been shown in \cite{R4621}
that the new physics effects shift the zero value of the forward--backward 
asymmetry for the $B \rar K^\ast \ell^+ \ell^-$ decay. Therefore we will study 
the sensitivity of the forward--backward asymmetry to the new Wilson 
coefficients. The normalized forward--backward asymmetry is defined as 
\bea
\label{e14}
{\cal A}_{FB} = \frac{\ds{\int_0^1\frac{d\Gamma}{dsdz}}\,dz -
\ds{\int_{-1}^0\frac{d\Gamma}{dsdz}}\,dz}
{\ds{\int_0^1\frac{d\Gamma}{dsdz}}\,dz +
\ds{\int_{-1}^0\frac{d\Gamma}{dsdz}}\,dz}~.
\eea

It is well known that ${\cal A}_{FB}$ is parity--odd but CP--even quantity,
which depends on the chirality of the lepton and quark currents. In order to
obtain $z=\cos \theta$ dependence, the differential decay width should
contain multiplication of such terms which transform even and odd under 
parity, respectively.  

\section{Numerical analysis}

In this section we will study the sensitivity of tee branching ratio and
lepton forward--backward asymmetry to the new Wilson coefficients. The main
input parameters in calculating the above--mentioned quantities are the
form factors.  Since there exists no exact calculation of the form factors
of the $\Lambda_b \rar \Lambda$ transition, we will use the form factors
derived from QCD sum rules in framework of the heavy quark effective theory,
which reduces the number of lots of form factors into two (see for example
\cite{R4629}). The $q^2$ dependence of these form factors can be represented
in terms of the three parameters as
\bea
F(q^2) = \frac{F(0)}{\ds 1-a_F\,\frac{q^2}{m_{\Lambda_b}^2} + b_F \left
    ( \frac{q^2}{m_{\Lambda_b}^2} \right)^2}~, \nnb
\eea
where parameters $F_i(0),~a$ and $b$ are listed in table 1 (see
\cite{R4630}) 
\begin{table}[h]    
\renewcommand{\arraystretch}{1.5} 
\addtolength{\arraycolsep}{3pt}  
$$
\begin{array}{|l|ccc|}
\hline
& F(0) & a_F & b_F \\ \hline
F_1 &
\phantom{-}0.462 & -0.0182 & -0.000176 \\
F_2 &
-0.077 & -0.0685 &\phantom{-}0.00146 \\ \hline
\end{array}
$$
\caption{Transition form factors for $\Lambda_b \rar \Lambda \ell^+ \ell^-$ 
decay in a three-parameter fit, where the
radiative corrections to the leading twist contribution and SU(3) breaking
effects are taken into account.}
\renewcommand{\arraystretch}{1}
\addtolength{\arraycolsep}{-3pt}
\end{table}

The values of other input parameters which appear in the expressions of the
branching ratio and forward--backward asymmetry are:
$m_b=4.8~GeV,~m_{\Lambda_b}=5.64~GeV,~m_\Lambda=1.116~GeV,~m_c=1.4~GeV$.
Contribution of new physics effects are contained in the new Wilson
coefficients (see Eq. (\ref{e1}). To the leading log approximation the 
values of the Wilson coefficients are
$C_7^{eff}=-0.313,~C_9^{eff}=4.344$ and $C_{10}^{eff}=-4.669$
\cite{R4614}.The value of the Wilson coefficient $C_9^{eff}$ used in the
numerical analysis corresponds only to short distance contribution. In
addition to this contribution $C_9^{eff}$ receives also long distance
contributions from the real $\bar c c$ intermediate states, i.e., from the
$J/\psi$ family. In the present work we do not take into consideration such 
contributions. In order to estimate the branching ratio and lepton
forward--backward asymmetry we need the values of the new Wilson coefficients
which describe new physics beyond the SM. In this work we will vary all new
Wilson coefficients within the range $-\vel C_{10} \ver \le C_X \le \vel
C_{10} \ver$. The experimental bounds on the branching ratio of the 
$B\rar K^\ast \mu^+\mu^-$ and $B_s\rar \mu^+\mu^-$ decays \cite{R4631}
suggests that this is the right order of magnitude range for the vector and
scalar Wilson coefficients. We assume that all new Wilson coefficients are
real, i.e., we do not introduce any new phase in addition to the one present
in the SM. 

Let us first study the dependence of the branching ratio for the $\Lambda_b
\rar \Lambda \ell^+ \ell^-$ decay on the new Wilson coefficients. In Figs.
(1--4) and (5--8) we present the dependence of the branching ratio for the
$\Lambda_b\rar \Lambda \mu^+ \mu^-$ ($\Lambda_b\rar \Lambda \tau^+
\tau^-$) decay on $C_{LL},~C_{LR},~C_{RR},~C_{RL},~C_{LRLR},~C_T$ and
$C_{TE}$, respectively. One can easily see from these figures that the
branching ratio is strongly dependent on $C_{LL}$ and the tensor interaction
coefficients $C_T$ and $C_{TE}$, while it is weakly dependent on the
remaining vector interaction couplings $C_{LR}$, $C_{RR}$ and $C_{RL}$ and 
the scalar coupling $C_{LRLR}$.
It should be noted that similar behavior takes place for the other scalar
interaction coefficients. Also, we observe from these figures that when
$C_{LL}>0$ ($C_{LL}<0$) contribution of the new Wilson coefficients to the SM
result is constructive (destructive). The situation is opposite for the
coefficient $C_{LR}$, i.e., it is constructive (destructive) when $C_{LR}<0$ 
($C_{LR}>0$). These behaviors can be explained as follows. We see from Eq.
(\ref{e2}) that $C_{LL}^{tot}=C_9^{eff}-C_{10}+C_{LL}$ and
$C_{LR}^{tot}=C_9^{eff}+C_{10}+C_{LR}$. Since $C_9^{eff}=4.344$ (short
distance) and $C_{10}=-4.669$ in the SM, contributions of $C_{LL}$ and
$C_{LR}$ are constructive (destructive) when $C_{LL}>0$ ($C_{LL}<0$) and
$C_{LR}<0$ ($C_{LR}>0$). 

We observe from Fig. (4) that the branching ratio is strongly
dependent on the tensor interaction. 

For the $\Lambda_b \rar \Lambda \tau^+ \tau^-$ decay the situation is
analogous to the $\Lambda_b \rar \Lambda \mu^+ \mu^-$ decay with a slight
difference. Contribution coming from different type vector interactions
becomes comparable. This fact can be explained by the fact that the terms
proportional to $\sim (1-v^2)$, which are very small in the $\mu$ case,
contribute more in the $\tau$ case. 

At this point we would like to remind that, similar dependence on the new
Wilson coefficients occurs for the $B \rar K^\ast \ell^+\ell^-$ decay. 

In Figs. (9)--(16) we present the dependence of the lepton forward--backward
asymmetry on the new Wilson coefficients for the $\Lambda_b \rar \Lambda
\mu^+ \mu^-$ and $\Lambda_b \rar \Lambda \tau^+ \tau^-$ decays. We observe
from Figs. (9)--(12) that, for the
$\Lambda_b \rar \Lambda \mu^+ \mu^-$ case the lepton forward--backward
asymmetry is more sensitive to the coefficients $C_{LL}$ and $C_{LR}$
and weakly depends on rest of the Wilson coefficients. It follows 
from these figures that when $C_{LL}$ is positive (negative), the zero point
of the forward--backward asymmetry is shifted to the left (right) from its
corresponding SM value. For all values of the coefficients
$C_{RR}$ and $C_{RL}$ the zero position of the forward--backward
asymmetry is shifted right and left with respect to its SM value,
respectively. It is observed in \cite{R4630} that, the zero
position of the dilepton forward--backward asymmetry in the $\Lambda_b \rar
\Lambda \ell^+\ell^-$ decay parametrically has very little dependence on
the form factors. Therefore the shift of zero position can be attributed to
the existence of new physics.  

So, in view of all these observations we can say that, determination of the
zero point of the forward--backward asymmetry can give us essential
information, not only about the existence of new physics, but also about the
sign of the new Wilson coefficients. 

From Figs. (13)--(16) we arrive at the following conclusion for the
$\Lambda_b \rar \Lambda \tau^+ \tau^-$ decay. Except tensor interaction
coefficients, the forward--backward asymmetry is negative for positive or 
negative values of the remaining ones. This situation is opposite to the 
$\Lambda_b \rar \Lambda \mu^+ \mu^-$ case. The value of the ${\cal A}_{FB}$
is more sensitive to the $C_{LRRL}$ and tensor interaction. Sign of the
${\cal A}_{FB}$ can give us unambiguous information about the sign of the
tensor interaction coefficients.

Obviously, investigation of polarization effects in the $\Lambda_b \rar
\Lambda \ell^+ \ell^-$ decay can provide us new information in addition to
the branching ratio and forward--backward asymmetry. We will consider this
question in one of our future works. 

In conclusion, a systematic analysis of the rare $\Lambda_b \rar \Lambda
\ell^+ \ell^-$ decay is presented. For the form factors describing the
$\Lambda_b \rar \Lambda$ transition we have used HQET predictions. The
sensitivity of the branching ratio and of the lepton forward--backward
asymmetry to the new Wilson coefficients is studied systematically. Analysis
of the zero position of the lepton forward--backward asymmetry determines
not only the magnitude but also the sign of the new Wilson coefficients for
the $\Lambda_b \rar \Lambda \mu^+ \mu^-$ decay. Sign of the
forward--backward asymmetry for the $\Lambda_b \rar \Lambda \tau^+ \tau^-$
decay can serve as a useful tool in determining the sign of the Wilson
coefficients.

\newpage

\bAPP{A}{Definition of the form factors}

As has already been noted, in describing the $\Lambda_b \rar \Lambda$
transition, the following matrix elements
\bea
&&\lla \Lambda \vel \bar s \gamma_\mu (1 \mp \gamma_5) b \ver \Lambda_b
\rra~,\nnb \\
&&\lla \Lambda \vel \bar s \sigma_{\mu\nu} (1 \mp \gamma_5) b \ver \Lambda_b \rra~,
\nnb \\
&&\lla \Lambda \vel \bar s (1 \mp \gamma_5) b \ver \Lambda_b \rra~.\nnb
\eea

These matrix elements are generally parametrized in the following way (here
we follow \cite{R4623}

\baeeq
\lla \Lambda \vel \bar s \gamma_\mu (1 \mp \gamma_5) b \ver \Lambda_b \rra  
\es \bar u_\Lambda \Big[ f_1 \gamma_\mu + i f_2 \sigma_{\mu\nu} q^\nu + f_3  
q_\mu \Big] u_{\Lambda_b}\\
\lla \Lambda \vel \bar s \gamma_\mu \gamma_5 b \ver \Lambda_b \rra
\es \bar u_\Lambda \Big[ g_1 \gamma_\mu \gamma_5 + i g_2 \sigma_{\mu\nu}
\gamma_5 q^\nu + g_3 q_\mu \gamma_5\Big] u_{\Lambda_b} \\
\lla \Lambda \vel \bar s \sigma_{\mu\nu} b \ver \Lambda_b \rra
\es \bar u_\Lambda \Big[ f_T \sigma_{\mu\nu} - i f_T^V \ga \gamma_\mu q^\nu -
\gamma_\nu q^\mu \dr - i f_T^S \ga P_\mu q^\nu - P_\nu q^\mu \dr \Big]
u_{\Lambda_b}\\
\lla \Lambda \vel \bar s \sigma_{\mu\nu} \gamma_5 b \ver \Lambda_b \rra
\es \bar u_\Lambda \Big[ g_T \sigma_{\mu\nu} - i g_T^V \ga \gamma_\mu q^\nu -
\gamma_\nu q^\mu \dr - i g_T^S \ga P_\mu q^\nu - P_\nu q^\mu \dr \Big]
\gamma_5 u_{\Lambda_b}
\eaeeq

The form factors of the magnetic dipole operators are defined as 
\baeeq
\lla \Lambda \vel \bar s i \sigma_{\mu\nu} q^\nu  b \ver \Lambda_b \rra
\es \bar u_\Lambda \Big[ f_1^T \gamma_\mu + i f_2^T \sigma_{\mu\nu} q^\nu
+ f_3^T q_\mu \Big] u_{\Lambda_b}\cr
\lla \Lambda \vel \bar s i \sigma_{\mu\nu}\gamma_5  q^\nu  b \ver \Lambda_b \rra
\es \bar u_\Lambda \Big[ g_1^T \gamma_\mu \gamma_5 + i g_2^T \sigma_{\mu\nu}
\gamma_5 q^\nu + g_3^T q_\mu \gamma_5\Big] u_{\Lambda_b}
\eaeeq 
\eAPP 
Multiplying (A3) and (A4) by $i q^\nu$ and comparing wit (A5) and (A6),
respectively, one can easily obtain the following relations
\baeeq
f_2^T \es f_T - f_T^S q^2~,\crcr
f_1^T \es \Big[ f_T^V + f_T^S \ga m_{\Lambda_b} + m_\Lambda\dr \Big] 
q^2~,\nnb \\
f_1^T \es - \frac{q^2}{m_{\Lambda_b} - m_\Lambda} f_3^T~,\nnb \\
g_2^T \es g_T - g_T^S q^2~,\\
g_1^T \es \Big[ g_T^V - g_T^S \ga m_{\Lambda_b} - m_\Lambda\dr \Big]
q^2~,\nnb \\
g_1^T \es \frac{q^2}{m_{\Lambda_b} + m_\Lambda} g_3^T~.\nnb
\eaeeq 

The matrix element of the scalar (pseudoscalar) operators $\bar s b$ and
$\bar s\gamma_5 b$ can be obtained from (A1) and (A2) by multiplying both
sides to $q^\mu$ and using equation of motion. Neglecting the mass of the
strange quark, we get
\baeeq
\lla \Lambda \vel \bar s b \ver \Lambda_b \rra \es \frac{1}{m_b} 
\bar u_\Lambda \Big[ f_1 \ga m_{\Lambda_b} - m_\Lambda \dr + f_3 q^2
\Big] u_{\Lambda_b}~,\\
\lla \Lambda \vel \bar s \gamma_5 b \ver \Lambda_b \rra \es \frac{1}{m_b} 
\bar u_\Lambda \Big[ g_1 \ga m_{\Lambda_b} + m_\Lambda\dr \gamma_5 - g_3 q^2
\gamma_5 \Big] u_{\Lambda_b}~.
\eaeeq

Using these definitions of the form factors and effective Hamiltonian in Eq.
(\ref{e1}), we get the following forms of the functions
$A_i,~B_i,~D_i,~E_i,~N_j$ and $H_j~,(i=1,2,3;~j=1,2)$ entering the matrix
element of the $\Lambda_b \rar \Lambda \ell^+ \ell^-$ decay:
\baeeq
A_1 \es \frac{1}{q^2}\ga f_1^T-g_1^T \dr C_{SL} + \frac{1}{q^2}\ga
f_1^T-g_1^T \dr C_{BR} + \frac{1}{2}\ga f_1-g_1 \dr \ga C_{LL}^{tot} +
C_{LR}^{tot} \dr \nnb \\
\ar \frac{1}{2}\ga f_1+g_1 \dr \ga C_{RL} + C_{RR} \dr~,\nnb \\
A_2 \es A_1 \ga 1 \rar 2 \dr ~,\nnb \\
A_3 \es A_1 \ga 1 \rar 3 \dr ~,\nnb \\
B_1 \es A_1 \ga g_1 \rar - g_1;~g_1^T \rar - g_1^T \dr ~,\nnb \\
B_2 \es B_1 \ga 1 \rar 2 \dr ~,\nnb \\
B_3 \es B_1 \ga 1 \rar 3 \dr ~,\nnb \\
D_1 \es \frac{1}{2} \ga C_{RR} - C_{RL} \dr \ga f_1+g_1 \dr +
\frac{1}{2} \ga C_{LR}^{tot} - C_{LL}^{tot} \dr \ga f_1-g_1 \dr~,\nnb \\
D_2 \es D_1 \ga 1 \rar 2 \dr ~, \\
D_3 \es D_1 \ga 1 \rar 3 \dr ~,\nnb \\
E_1 \es D_1 \ga g_1 \rar - g_1 \dr ~,\nnb \\
E_2 \es E_1 \ga 1 \rar 2 \dr ~,\nnb \\
E_3 \es E_1 \ga 1 \rar 3 \dr ~,\nnb \\
N_1 \es \frac{1}{m_b} \Big( f_1 \ga m_{\Lambda_b} - m_\Lambda\dr + f_3 q^2
\Big) \Big( C_{LRLR} + C_{RLLR} + C_{LRRL} + C_{RLRL} \Big)~,\nnb \\
N_2 \es N_1 \ga C_{LRRL} \rar - C_{LRRL};~C_{RLRL} \rar - C_{RLRL} \dr~,\nnb \\
H_1 \es \frac{1}{m_b} \Big( g_1 \ga m_{\Lambda_b} + m_\Lambda\dr - g_3 q^2  
\Big) \Big( C_{LRLR} - C_{RLLR} + C_{LRRL} - C_{RLRL} \Big)~,\nnb \\
H_2 \es H_1 \ga C_{LRRL} \rar - C_{LRRL};~C_{RLRL} \rar - C_{RLRL} \dr~.\nnb
\eaeeq

\eAPP

\newpage

\bAPP{B}{Double differential rate}

The explicit form of the expressions ${\cal T}_0(s)$, ${\cal T}_1(s)$
and ${\cal T}_2(s)$ are as follows:
\baeeq
\label{e9}
{\cal T}_0(s) \es - 2048 \lambda m_\ell^2 m_{\Lambda_b}^4 \vel C_T 
\ver^2 \mbox{\rm Re}[f_T^\ast f_T^S]\nnb \\  
\ar 384 m_\ell m_{\Lambda_b}^3 \Big\{
(1 + \sqrt{r}) (1 - 2 \sqrt{r} + r - s) 
\mbox{\rm Re}[(A_1+B_1)^\ast C_T f_T] \nnb \\
\ar 2 (1 - \sqrt{r}) (1 + 2 \sqrt{r} + r - s) 
\mbox{\rm Re}[(A_1-B_1)^\ast C_{TE} f_T] \Big\} \nnb \\
\ar 32 m_\ell^2 m_{\Lambda_b}^4 s (1+r-s) \ga \vel D_3 \ver^2 +
\vel E_3 \ver^2 \dr \nnb \\
\ar 4 m_{\Lambda_b}^4 s (1-2\sqrt{r}+r-s) \Big( 4 m_\ell \, 
\mbox{\rm Re} [\ga D_3 - E_3 \dr^\ast H_2] + \vel H_2 \ver^2\Big) \nnb \\  
\ar 64 m_\ell^2 m_{\Lambda_b}^3 (1-r-s) \mbox{\rm Re} [D_1^\ast E_3 + D_3
E_1^\ast] \nnb \\
\ar 256 m_\ell m_{\Lambda_b}^4 (1+2\sqrt{r}+r-s) 
(2 - 4 \sqrt{r}+2 r+s) \mbox{\rm Re} [A_2^\ast C_{TE} f_T]\nnb \\
\ek 128 \lambda m_\ell m_{\Lambda_b}^5 \Big\{ 
(1+\sqrt{r}) \Big( \mbox{\rm Re} [(A_1+B_1)^\ast C_T f_T^S] - 16 m_\ell 
\vel C_T \ver^2
\mbox{\rm Re} [{f_T^S}^\ast f_T^V] \Big)\nnb \\
\ek m_{\Lambda_b} s \, 
\mbox{\rm Re} [(A_2 + B_2)^\ast C_T f_T^S] \Big\} \nnb \\
\ar 64 m_{\Lambda_b}^2 \sqrt{r} (6 m_\ell^2 - m_{\Lambda_b}^2 s)
{\rm Re} [D_1^\ast E_1] \nnb \\
\ek 128 m_\ell m_{\Lambda_b}^4 \Big[(1-r)^2 + (1-6\sqrt{r} + 
r) s - 2 s^2 \Big] \mbox{\rm Re} [(A_1+B_1)^\ast C_T f_T^V]  \nnb \\  
\ar 64 m_\ell^2 m_{\Lambda_b}^3 \sqrt{r} 
\Big( 2 m_{\Lambda_b} s {\rm Re} [D_3^\ast E_3] + (1 - r + s) 
{\rm Re} [D_1^\ast D_3 + E_1^\ast E_3]\Big) \nnb \\
\ek 128 m_\ell m_{\Lambda_b}^4
\Big\{ 2 (1+2\sqrt{r}+r-s) (2-4\sqrt{r}+2 r+s)
\mbox{\rm Re}[B_2^\ast C_{TE} f_T] \nnb \\
\ar (1-2\sqrt{r}+r-s) (2+4\sqrt{r}+2 r+s)
\mbox{\rm Re}[B_2^\ast C_T f_T] \Big\}\nnb \\
\ar 32 m_{\Lambda_b}^2 (2 m_\ell^2 + m_{\Lambda_b}^2 s)
\Big\{ (1 - r + s) m_{\Lambda_b} \sqrt{r} \,
{\rm Re} [A_1^\ast A_2 + B_1^\ast B_2] \\
\ek m_{\Lambda_b} (1 - r - s)  {\rm Re} [A_1^\ast B_2 + A_2^\ast B_1] - 
2 \sqrt{r} \Big( {\rm Re} [A_1^\ast B_1] + m_{\Lambda_b}^2 s 
{\rm Re} [A_2^\ast B_2] \Big) \Big\} \nnb \\
\ar 8 m_{\Lambda_b}^2 \Big\{ 4 m_\ell^2 (1 + r - s) + 
m_{\Lambda_b}^2 \Big[(1-r)^2 - s^2 \Big] 
\Big\} \ga \vel A_1 \ver^2 +  \vel B_1 \ver^2 \dr \nnb \\ 
\ar 8 m_{\Lambda_b}^4 \Big\{ 4 m_\ell^2 \Big[ \lambda + 
(1 + r - s) s \Big] + 
m_{\Lambda_b}^2 s \Big[(1-r)^2 - s^2 \Big] 
\Big\} \ga \vel A_2 \ver^2 +  \vel B_2 \ver^2 \dr \nnb \\
\ek 8 m_{\Lambda_b}^2 \Big\{ 4 m_\ell^2 (1 + r - s) - 
m_{\Lambda_b}^2 \Big[(1-r)^2 - s^2 \Big] 
\Big\} \ga \vel D_1 \ver^2 +  \vel E_1 \ver^2 \dr \nnb \\
\ar 512 m_{\Lambda_b}^2 \vel f_T \ver^2 \Big\{
\Big[ 2 m_\ell^2 (1 - 6 \sqrt{r} + r - s) 
+m_{\Lambda_b}^2 [\lambda + (1 + r - s) s ] \Big] 
\vel C_T \ver ^2 \nnb \\
\ar 4 \Big[ 2 m_\ell^2 (1 + 6 \sqrt{r} + r - s) + m_{\Lambda_b}^2 
[\lambda + (1 + r - s) s ] \Big] \vel C_{TE} \ver ^2 \Big\} \nnb \\
\ar 8 m_{\Lambda_b}^5 s v^2 \Big\{
- 8 m_{\Lambda_b} s \sqrt{r}\, {\rm Re} [D_2^\ast E_2] +
4 (1 - r + s) \sqrt{r} \, {\rm Re} [D_1^\ast D_2+E_1^\ast E_2]\nnb \\
\ek 4 (1 - r - s) {\rm Re} [D_1^\ast E_2+D_2^\ast E_1] +
m_{\Lambda_b} \Big[(1-r)^2 -s^2\Big]   
\ga \vel D_2 \ver^2 + \vel E_2 \ver^2\dr \Big\}\nnb \\
\ar (1 + 2 \sqrt{r} + r - s)\Big\{
 1024 \lambda m_\ell^2 m_{\Lambda_b}^6 \vel C_T \ver^2 \vel f_T^S \ver^2 \nnb \\
\ar 16 m_\ell m_{\Lambda_b}^3 (1-\sqrt{r}) 
{\rm Re} [(D_1+E_1)^\ast F_2] \nnb \\
\ar 4 m_{\Lambda_b}^4 s \vel F_2 \ver^2 +
4 m_{\Lambda_b}^4 s \Big( 4 m_\ell \, {\rm Re} [(D_3+E_3)^\ast F_2]
+ v^2 \vel F_1 \ver^2 \Big) \Big\} \nnb \\
\ar (1 - 2 \sqrt{r} + r - s)\Big\{
- 128 m_\ell m_{\Lambda_b}^4 (2 + 4 \sqrt{r} + 2 r + s)
{\rm Re} [A_2^\ast C_T f_T] \nnb \\
\ar 512 m_{\Lambda_b}^3 (1 + \sqrt{r}) 
{\rm Re} [f_T^\ast f_T^V] \Big[ 8 m_\ell^2 \ga 2 \vel C_{TE} \ver^2 
- \vel C_T \ver^2 \dr - m_{\Lambda_b}^2 s \ga 4 \vel C_{TE} \ver^2 + \vel C_T
\ver^2 \dr \Big] \nnb \\
\ek 16 m_\ell m_{\Lambda_b}^3 (1 + \sqrt{r}) \Big[
{\rm Re} [(D_1-E_1)^\ast H_2] - 24 m_{\Lambda_b}^2 s \, {\rm Re}
[(A_2+B_2)^\ast C_T f_T^V  \Big] \nnb \\
\ar 4 m_{\Lambda_b}^4 s v^2 \vel H_1 \ver^2 \nnb \\
\ar 256 m_{\Lambda_b}^4 \vel f_T^V \ver^2 \Big(
 \Big[ m_{\Lambda_b}^2 s^2 + 4 m_\ell^2 
(1 + 2 \sqrt{r} + r + s) \Big] \vel C_T \ver^2 + 4  m_{\Lambda_b}^2 s^2 v^2 
\vel C_{TE} \ver^2 \Big) \Big\}~,
\nnb \\ \nnb \\
\label{e10}
{\cal T}_1(s) \es - 16 m_\ell m_{\Lambda_b}^3 v \sqrt{\lambda} \Big\{
(1-\sqrt{r}) {\rm Re} [(A_1-B_1)^\ast H_1]-
(1+\sqrt{r}) {\rm Re} [(A_1+B_1)^\ast F_1] \Big\} \nnb \\
\ek 384 m_\ell m_{\Lambda_b}^3 v \sqrt{\lambda} \Big\{
(1+\sqrt{r}) {\rm Re} [(D_1-E_1)^\ast C_T f_T] +
2 (1-\sqrt{r}) {\rm Re} [(D_1+E_1)^\ast C_{TE} f_T]\Big\} \nnb \\
\ek 256 m_\ell m_{\Lambda_b}^4 v \sqrt{\lambda} (1-r) 
\Big( {\rm Re} [(D_2-E_2)^\ast C_T f_T] -
2 {\rm Re} [(D_2+E_2)^\ast C_{TE} f_T] \Big) \nnb \\
\ar 256 m_\ell m_{\Lambda_b}^5  v \sqrt{\lambda} (1-\sqrt{r}) 
(1 + 2 \sqrt{r} + r - s) 
{\rm Re} [(D_1+E_1)^\ast C_{TE} f_T^S] \Big)  \nnb \\
\ar 128 m_\ell m_{\Lambda_b}^4 s v \sqrt{\lambda} 
\Big( {\rm Re} [(C_T-2 C_{TE})^\ast D_3 f_T^\ast] -
{\rm Re} [(C_T+2 C_{TE})^\ast E_3 f_T^\ast] \Big) \nnb \\
\ek 16 m_{\Lambda_b}^4 s v \sqrt{\lambda}
\Big\{ 2 {\rm Re} [A_1^\ast D_1] - 2 {\rm Re} [B_1^\ast E_1] -
4 {\rm Re} [(F_1+H_2)^\ast C_T f_T] \nnb \\
\ar 8 {\rm Re} [(F_2+H_1)^\ast C_{TE} f_T]
+ m_\ell {\rm Re} [(A_2+B_2)^\ast F_1] \Big\} \\
\ek 16 m_{\Lambda_b}^4 s v \sqrt{\lambda} \Big(
m_\ell {\rm Re} [(A_2-B_2)^\ast H_1]
+2 m_{\Lambda_b}{\rm Re} [B_1^\ast D_2 - B_2^\ast D_1 +
A_2^\ast E_1 - A_1^\ast E_2] \Big) \nnb \\
\ar 256 m_\ell m_{\Lambda_b}^5 s v \sqrt{\lambda} (1- \sqrt{r})
{\rm Re} [(D_2-E_2)^\ast C_T f_T^V] \nnb \\
\ar 64  m_{\Lambda_b}^5 s v \sqrt{\lambda} (1+ \sqrt{r})
\Big( - {\rm Re} [F_1^\ast C_T f_T^V] + 
2 {\rm Re} [F_2^\ast C_{TE} f_T^V]\nnb \\
\ar 4 m_\ell {\rm Re} [(D_3+E_3)^\ast C_{TE} f_T^V]\Big) \nnb \\
\ar 32 m_{\Lambda_b}^6 s v \sqrt{\lambda} (1- r)
{\rm Re} [A_2^\ast D_2 - B_2^\ast E_2] \nnb \\
\ar 32 m_{\Lambda_b}^5 s v \sqrt{\lambda} \sqrt{r}
{\rm Re} [A_2^\ast D_1 + A_1^\ast D_2 - B_2^\ast E_1 - B_1^\ast E_2] \nnb \\
\ar 64 m_{\Lambda_b}^6 s v \sqrt{\lambda} (1+2 \sqrt{r} + r -s)
\Big( - {\rm Re} [F_1^\ast C_T f_T^S] +
2 {\rm Re} [F_2^\ast C_{TE} f_T^S]\nnb \\
\ar 4 m_\ell {\rm Re} [(D_3+E_3)^\ast C_{TE} f_T^S]\Big) \nnb \\
\ar 256 m_\ell m_{\Lambda_b}^4 v \sqrt{\lambda}
 \Big\{ (1-r) {\rm Re} [(D_1+E_1)^\ast C_{TE} f_T^V]
+ s {\rm Re} [(D_1-E_1)^\ast C_T f_T^V] \Big\}~,
\nnb \\ \nnb \\
\label{e11}
{\cal T}_0(s) \es - 8 m_{\Lambda_b}^4 v^2 \lambda 
\ga \vel A_1 \ver^2 + \vel B_1 \ver^2 + \vel D_1 \ver^2 
+ \vel E_1 \ver^2 \dr \nnb \\
\ek 512 m_{\Lambda_b}^4 v^2 \lambda \Big[ \ga 4 \vel C_{TE} \ver^2 +
\vel C_T \ver^2 \dr \vel f_T \ver^2 \Big] \\
\ar 8 m_{\Lambda_b}^6 s v^2 \lambda 
\Big( \vel A_2 \ver^2 +
\vel B_2 \ver^2 + \vel D_2 \ver^2 + \vel E_2 \ver^2  \Big) \nnb  \\
\ek 256 m_{\Lambda_b}^6 s v^2 \lambda \ga 4 \vel C_{TE} \ver^2 + \vel C_T
\ver^2 \dr \Big\{ 2 \, {\rm Re} [f_T^\ast f_T^S] - 
\vel f_T^V + m_{\Lambda_b} (1+\sqrt{r}) f_T^S \ver^2 \nnb \\
\ar m_{\Lambda_b}^2 s \vel f_T^S \ver^2 \Big\}~. \nnb    
\eaeeq

\eAPP

\newpage

\section*{Figure captions}
{\bf Fig. (1)} The dependence of the branching ratio for the 
$\Lambda_b \rar \Lambda \mu^+ \mu^-$ decay on the new Wilson coefficients
$C_{LL}$ and $C_{LR}$.\\ \\
{\bf Fig. (2)} The same as in Fig. (1), but for the coefficients 
$C_{RR}$ and $C_{LR}$.\\ \\
{\bf Fig. (3)} The same as in Fig. (1), but for the coefficient 
$C_{LRRL}$.\\ \\
{\bf Fig. (4)} The same as in Fig. (1), but for the coefficients 
$C_T$ and $C_{TE}$, describing the tensor interactions.\\ \\
{\bf Fig. (5)} The dependence of the branching ratio for the 
$\Lambda_b \rar \Lambda \tau^+ \tau^-$ decay on the new Wilson coefficients 
$C_{LL}$ and $C_{LR}$.\\ \\
{\bf Fig. (6)} The same as in Fig. (5), but for the coefficients 
$C_{RR}$ and $C_{LR}$.\\ \\
{\bf Fig. (7)} The same as in Fig. (5), but for the coefficient  
$C_{LRRL}$.\\ \\
{\bf Fig. (8)} The same as in Fig. (5), but for the coefficients 
$C_T$ and $C_{TE}$ describing the tensor interactions.\\ \\
{\bf Fig. (9)} The dependence of the lepton forward--backward asymmetry 
for the  $\Lambda_b \rar \Lambda \mu^+ \mu^-$ decay on the new Wilson 
coefficients $C_{LL}$ and $C_{LR}$.\\ \\
{\bf Fig. (10)} The same as in Fig. (9), but for the coefficients
$C_{RR}$ and $C_{LR}$.\\ \\
{\bf Fig. (11)} The same as in Fig. (9), but for the coefficient
$C_{LRRL}$.\\ \\
{\bf Fig. (12)} The same as in Fig. (9), but for the coefficients
$C_T$ and $C_{TE}$, describing the tensor interactions.\\ \\
{\bf Fig. (13)} The dependence of the lepton forward--backward asymmetry 
for the  $\Lambda_b \rar \Lambda \tau^+ \tau^-$ decay on the new Wilson 
coefficients $C_{LL}$ and $C_{LR}$.\\ \\
{\bf Fig. (14)} The same as in Fig. (13), but for the coefficients
$C_{RR}$ and $C_{LR}$.\\ \\
{\bf Fig. (15)} The same as in Fig. (13), but for the coefficient
$C_{LRRL}$.\\ \\
{\bf Fig. (16)} The same as in Fig. (13), but for the coefficients
$C_T$ and $C_{TE}$, describing the tensor interactions.\\ \\

\newpage

\begin{figure}
\vskip 1.5 cm
    \includegraphics{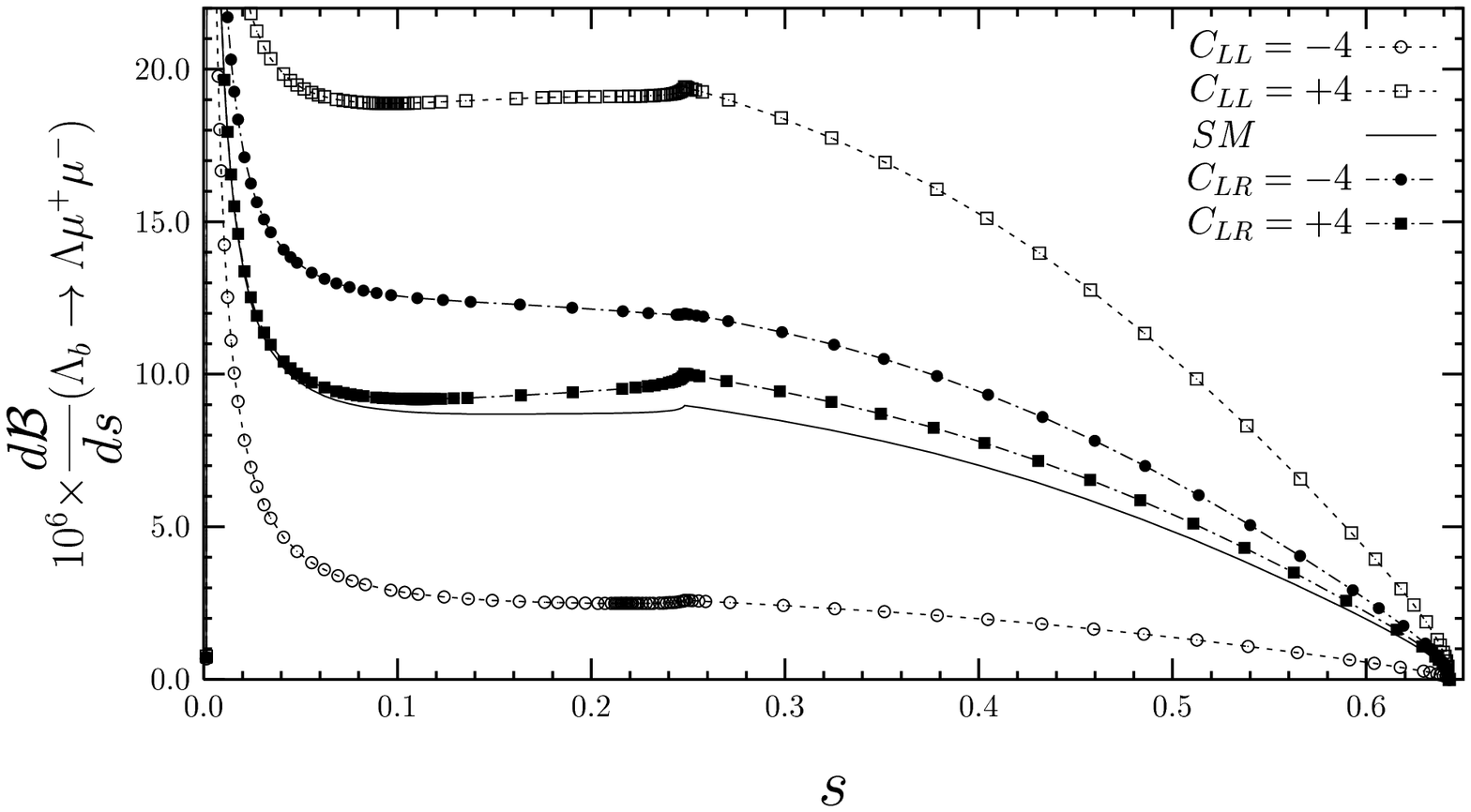}
\vskip 7.8cm
\caption{}
\end{figure}  

\begin{figure}   
\vskip 2.5 cm
    \includegraphics{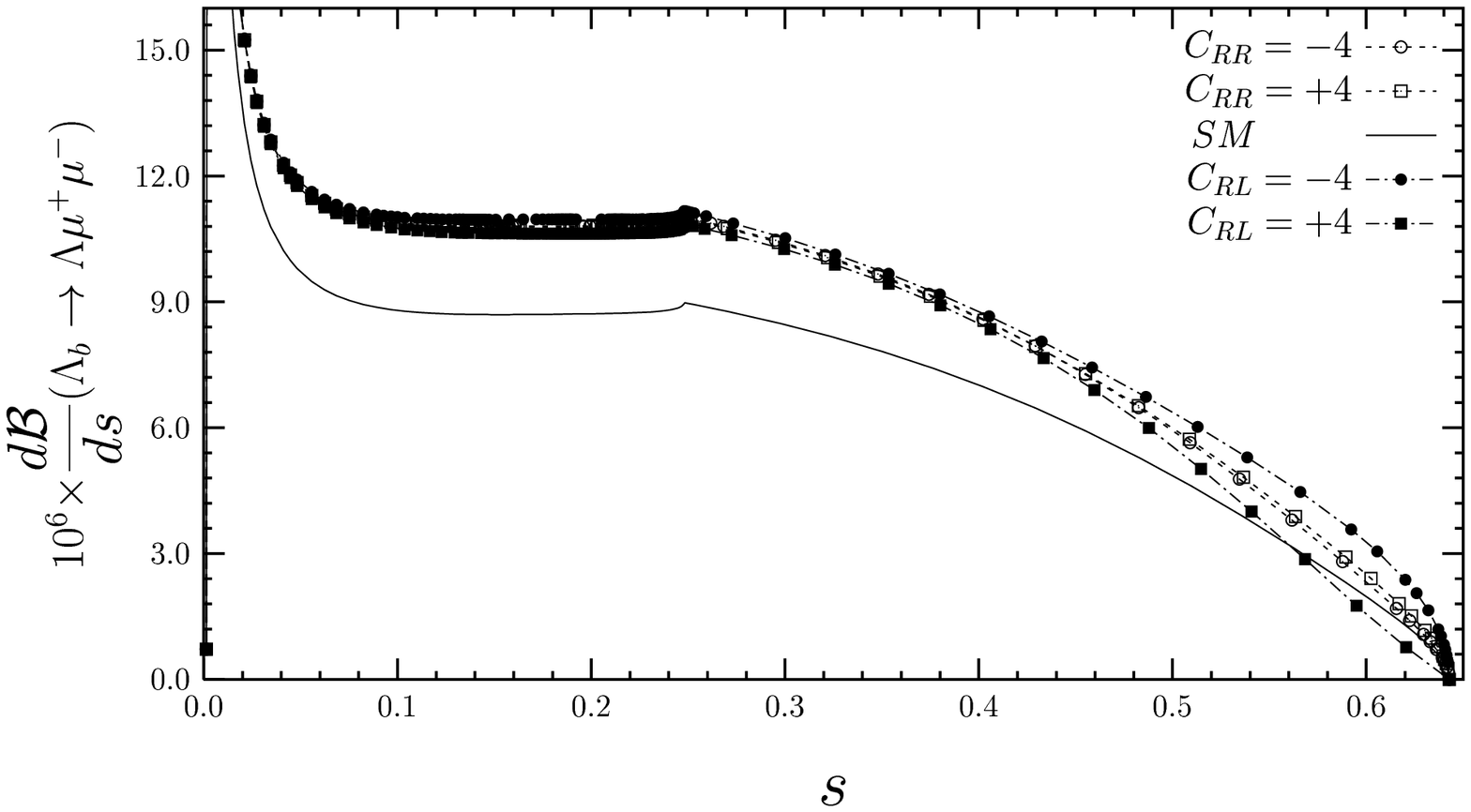}
\vskip 7.8 cm   
\caption{}
\end{figure}

\begin{figure}   
\vskip 1.5 cm
    \includegraphics{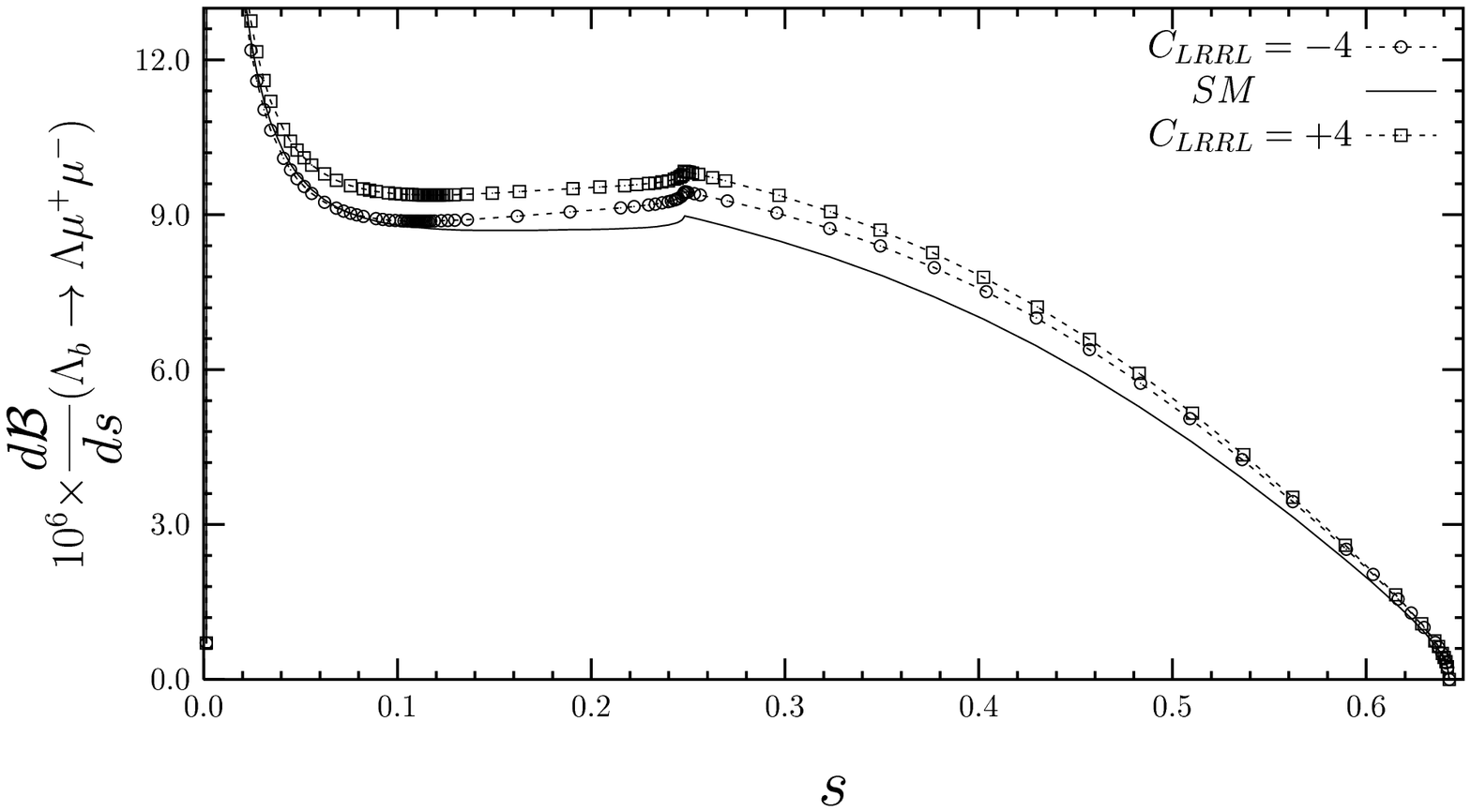}
\vskip 7.8cm
\caption{}
\end{figure}

\begin{figure}    
\vskip 2.5 cm
    \includegraphics{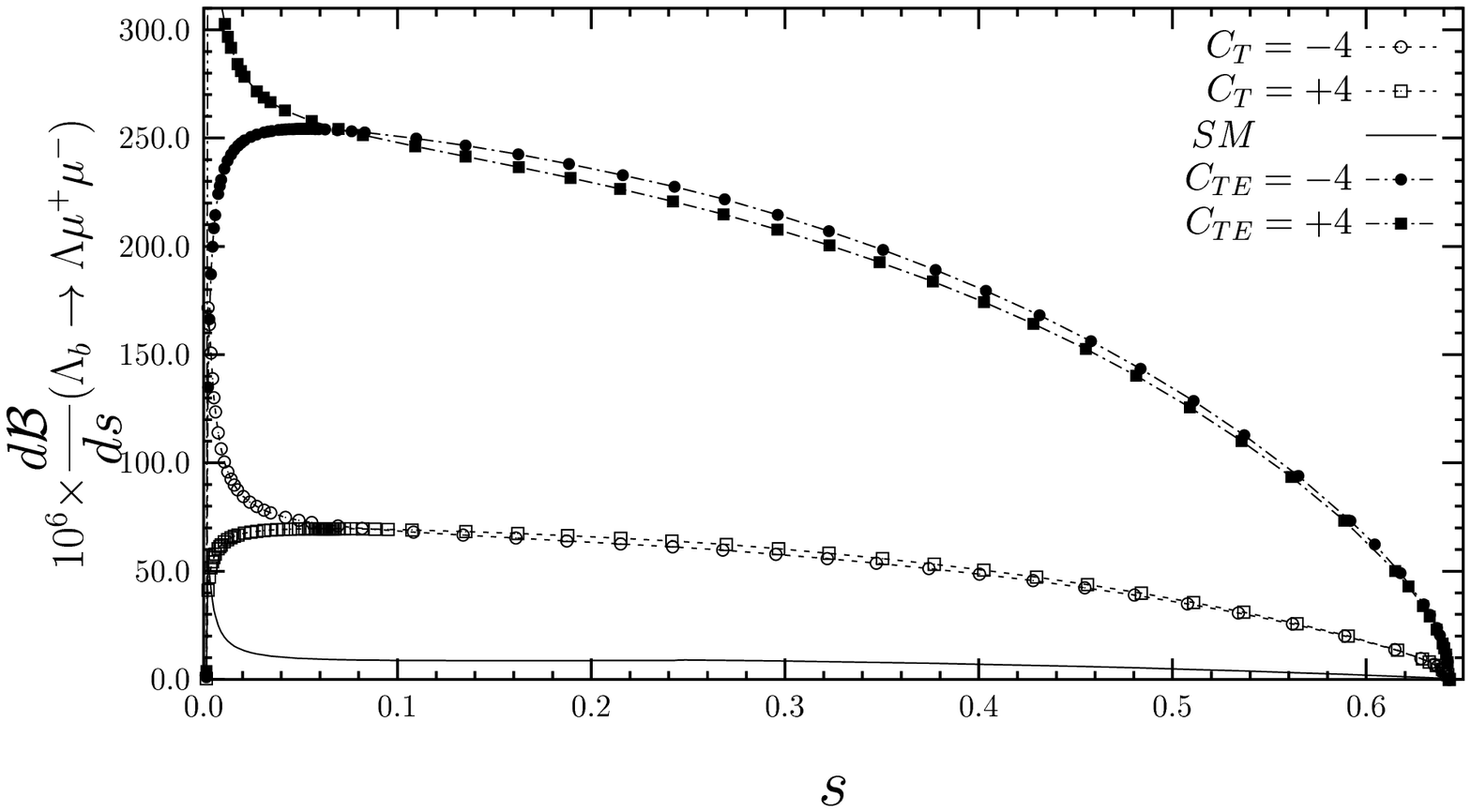}
\vskip 7.8 cm   
\caption{}
\end{figure}

\begin{figure}
\vskip 1.5 cm
    \includegraphics{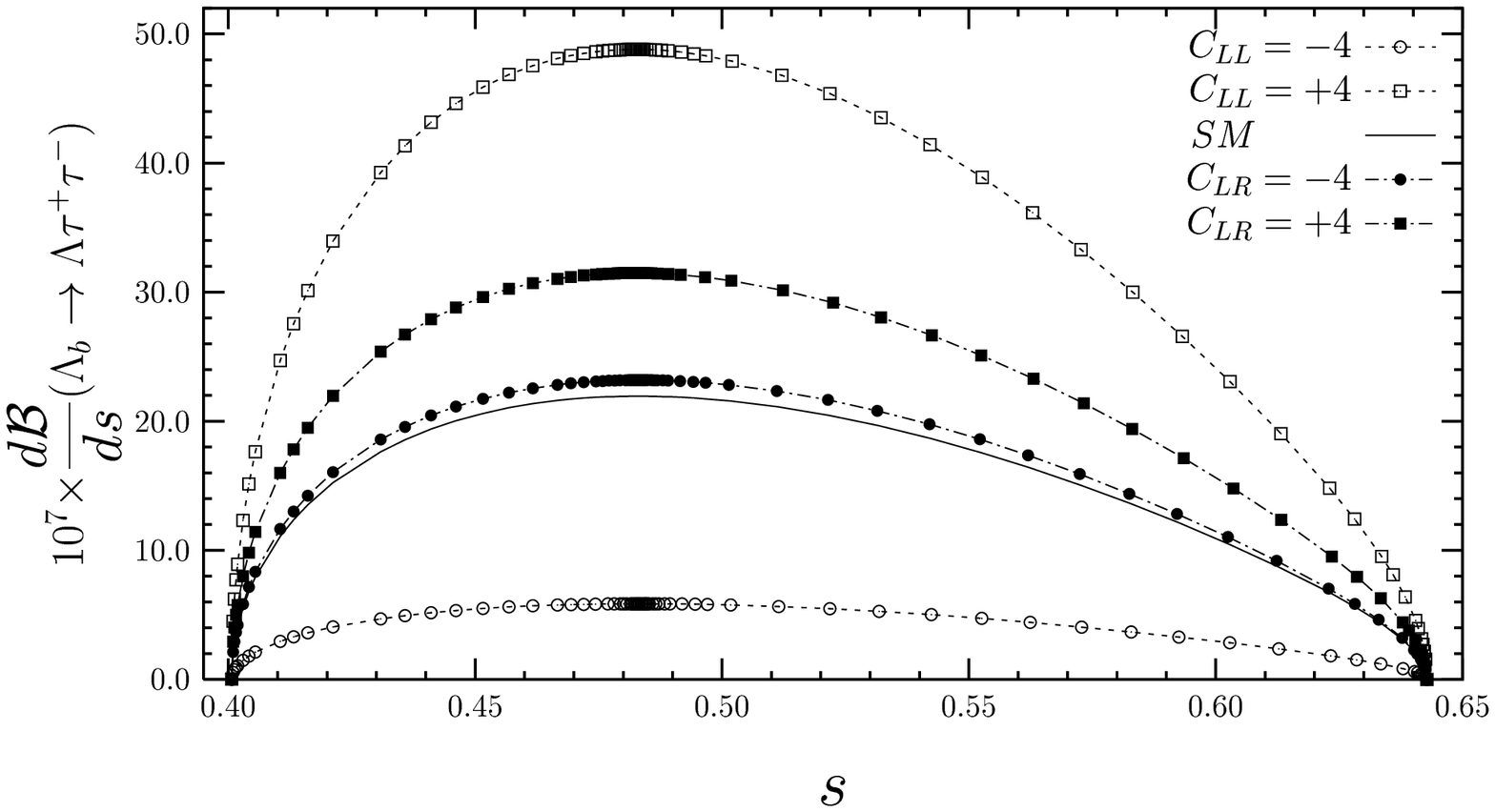}
\vskip 7.8cm
\caption{}
\end{figure}

\begin{figure}   
\vskip 2.5 cm
    \includegraphics{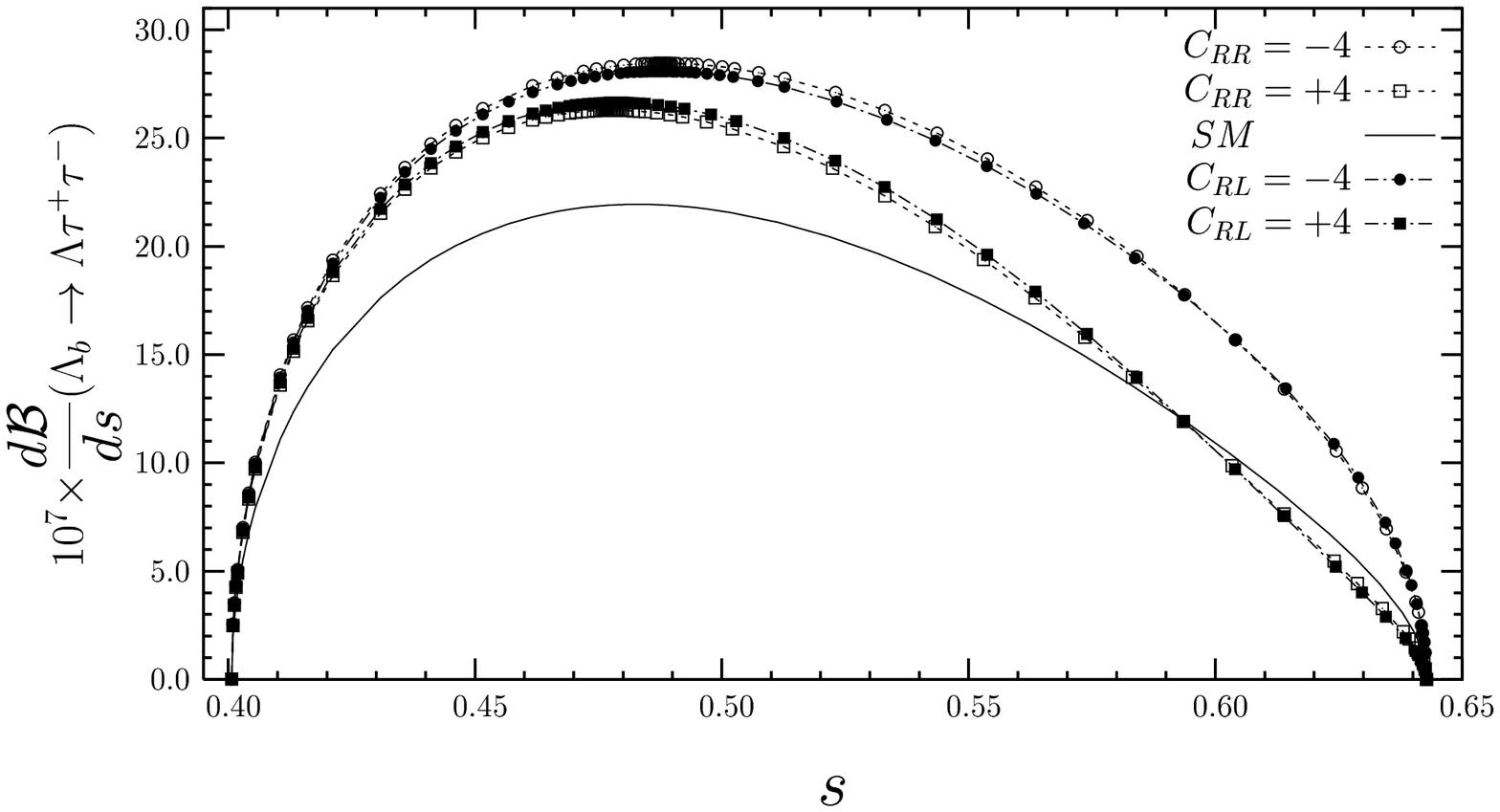}
\vskip 7.8 cm    
\caption{}
\end{figure}

\begin{figure}
\vskip 1.5 cm
    \includegraphics{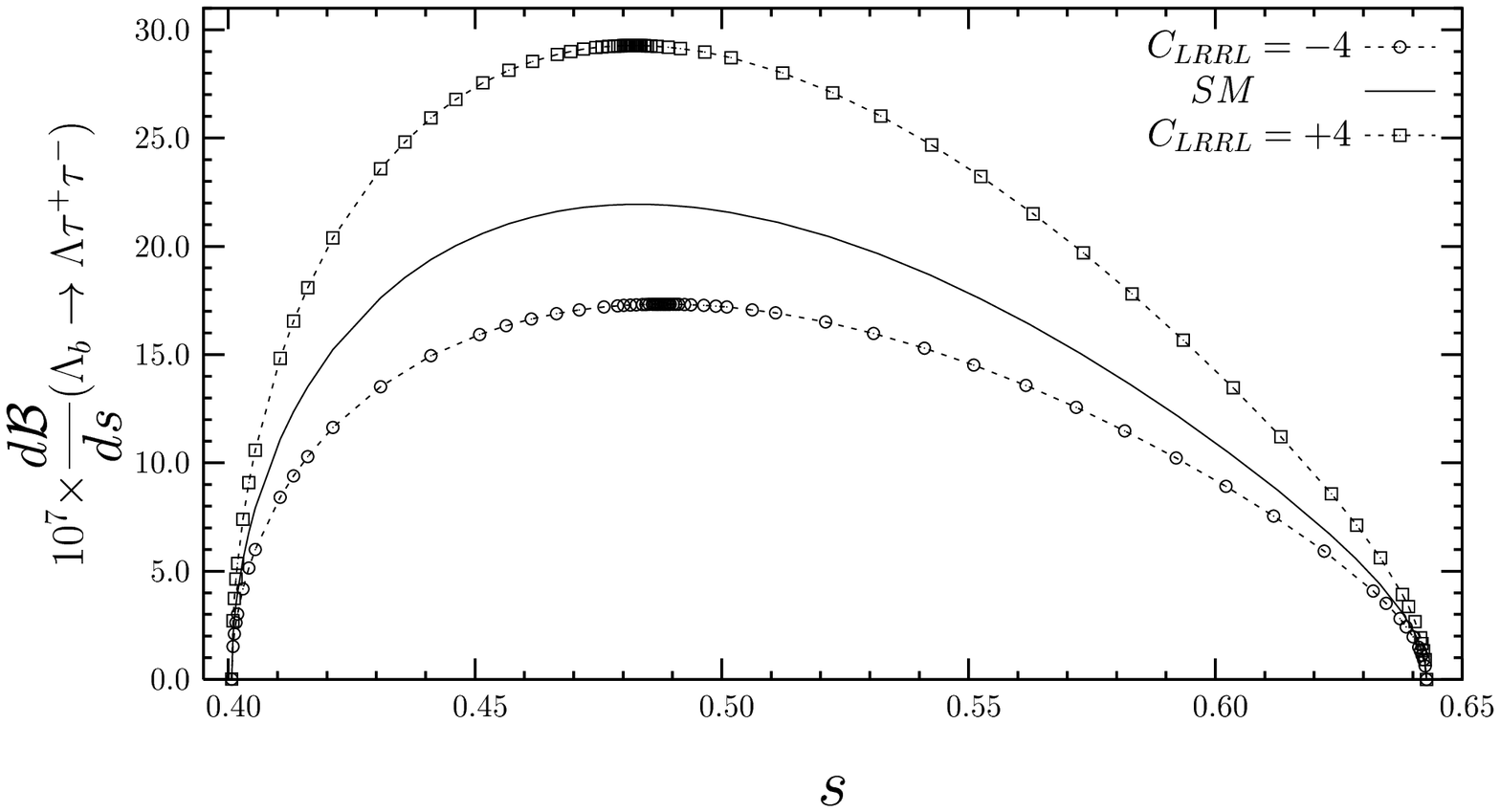}
\vskip 7.8cm
\caption{}
\end{figure}  

\begin{figure}   
\vskip 2.5 cm
    \includegraphics{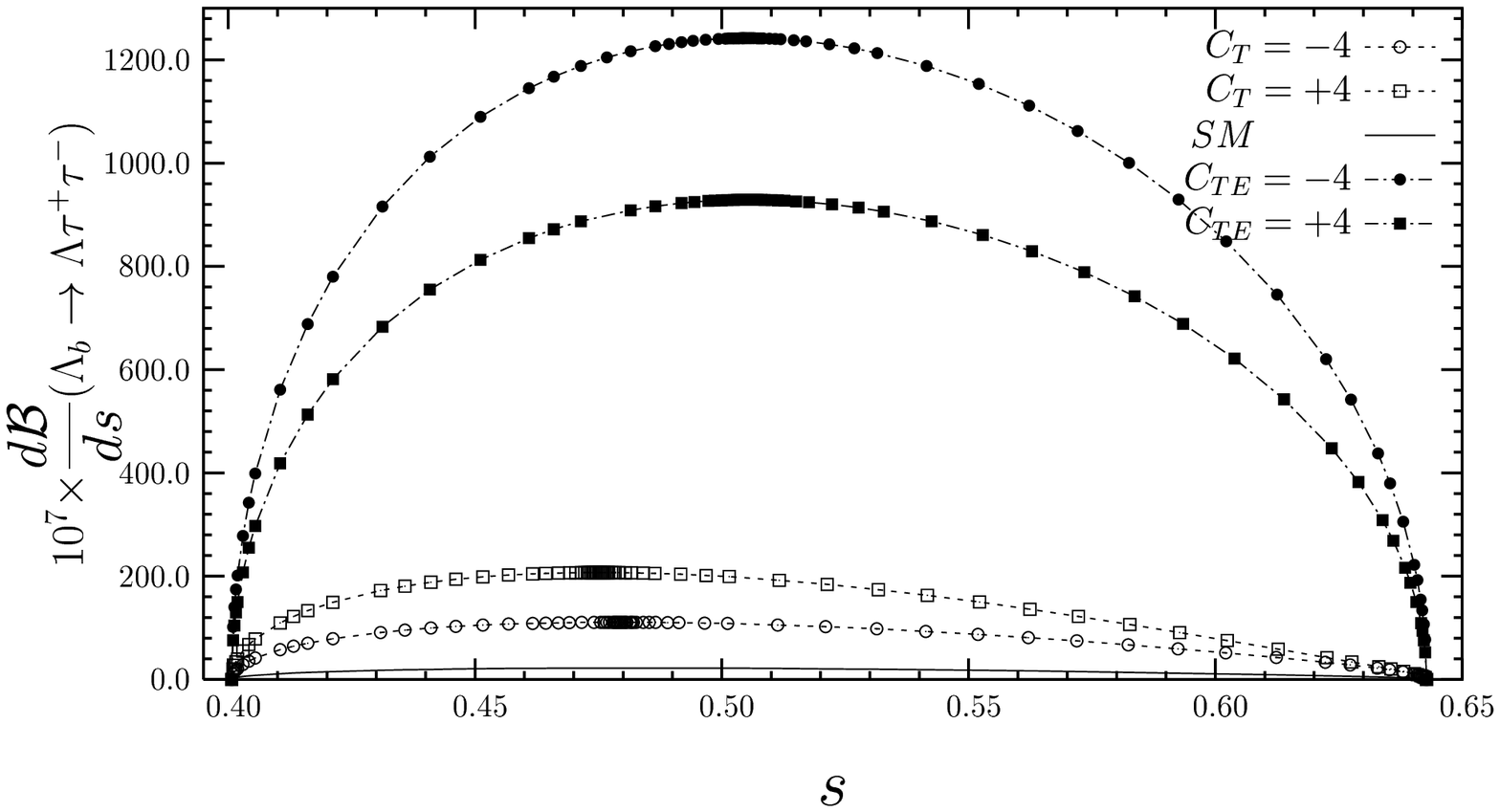}
\vskip 7.8 cm   
\caption{}
\end{figure}

\begin{figure}   
\vskip 1.5 cm
    \includegraphics{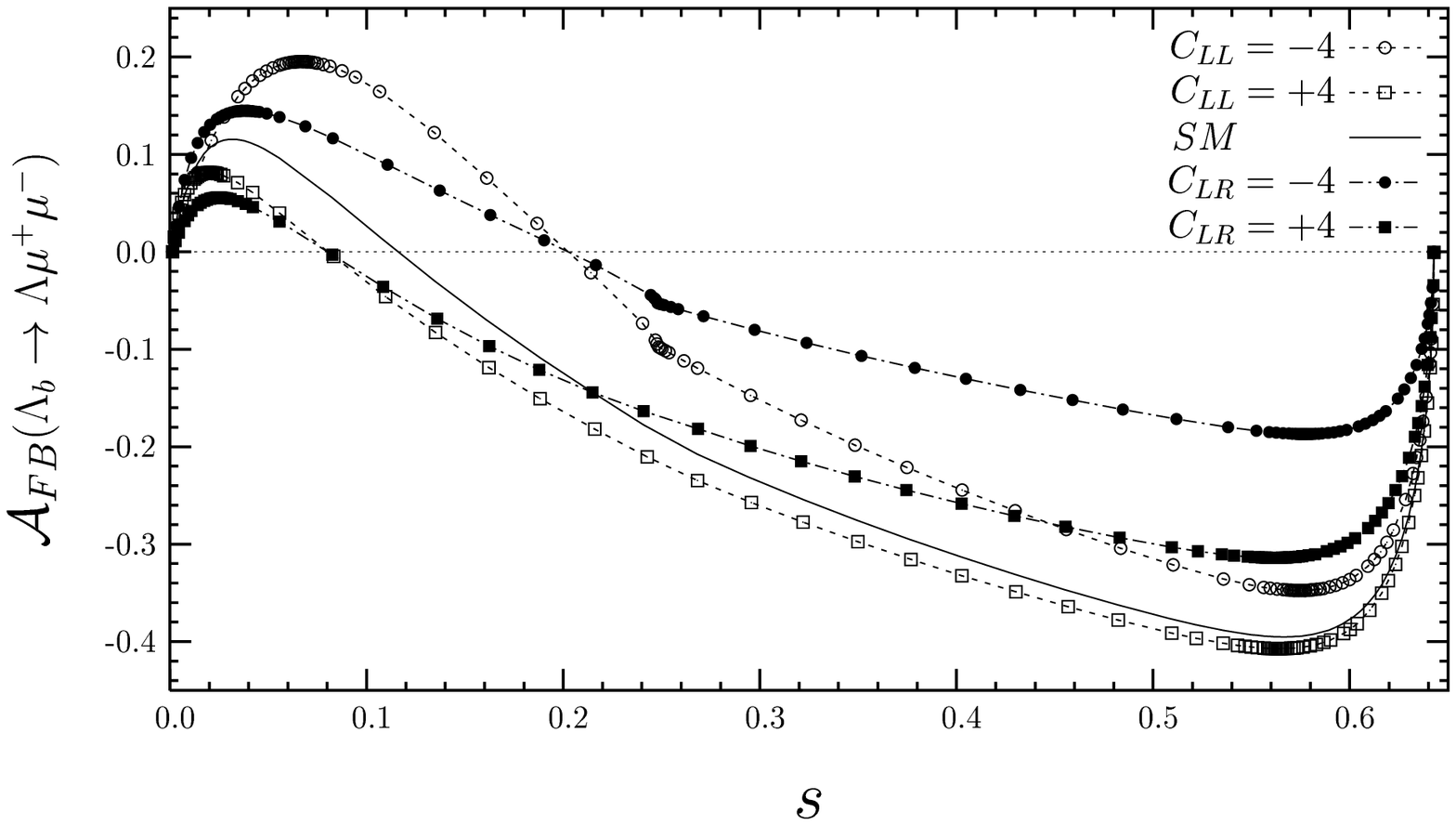}
\vskip 7.8cm
\caption{}
\end{figure}

\begin{figure}    
\vskip 2.5 cm
    \includegraphics{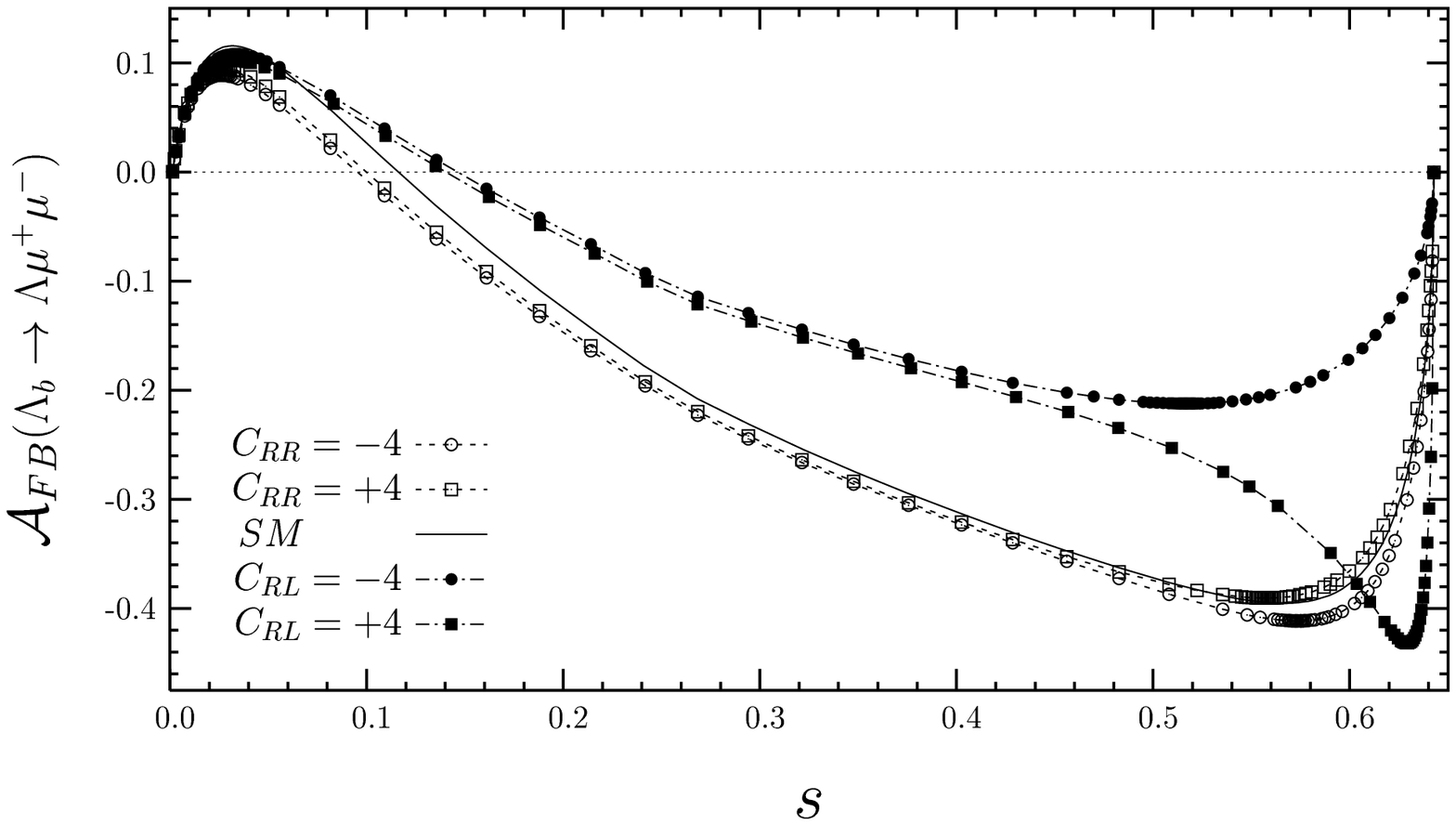}
\vskip 7.8 cm   
\caption{}
\end{figure}

\begin{figure}
\vskip 1.5 cm
    \includegraphics{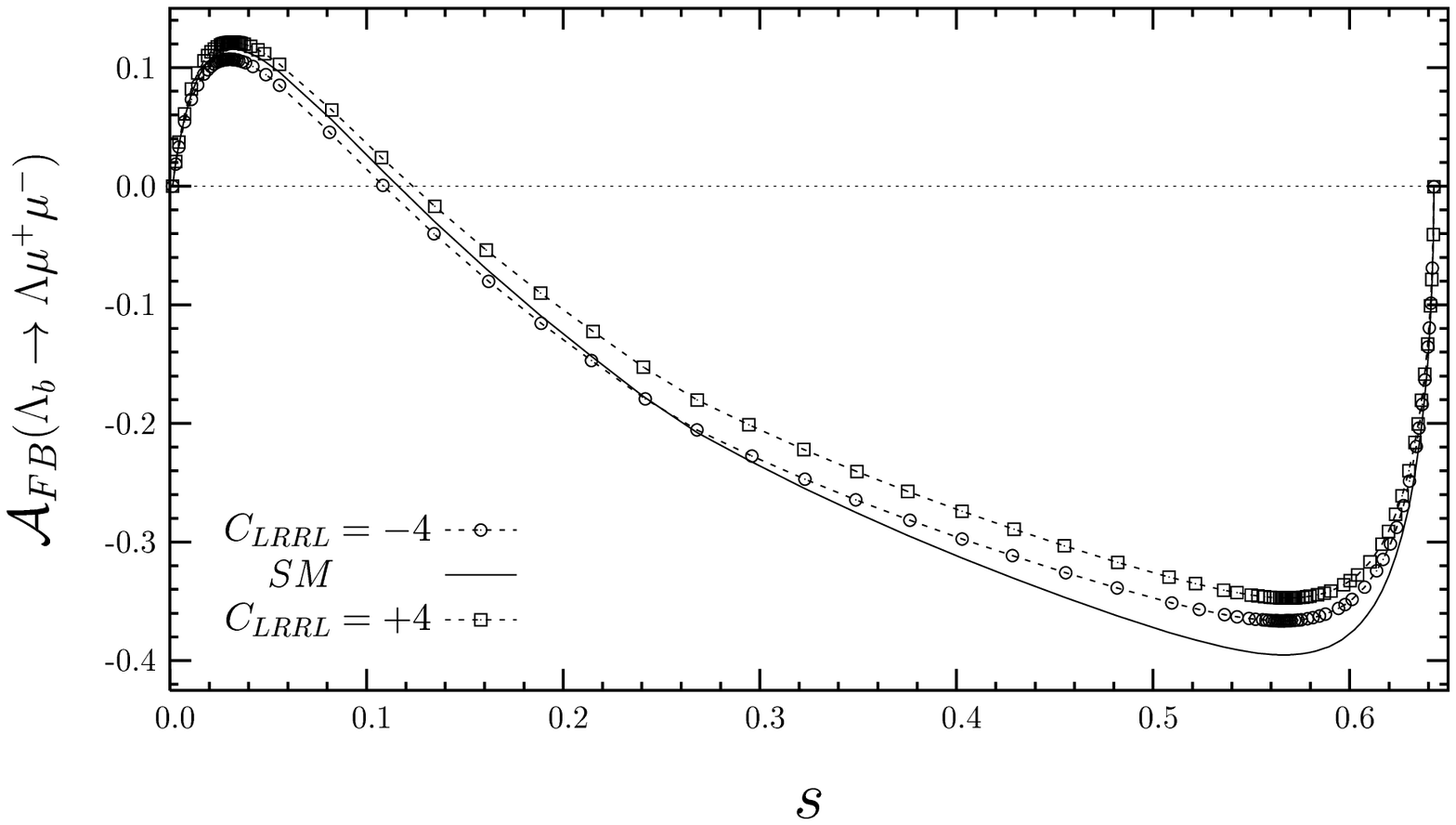}
\vskip 7.8cm
\caption{}
\end{figure}

\begin{figure}   
\vskip 2.5 cm
    \includegraphics{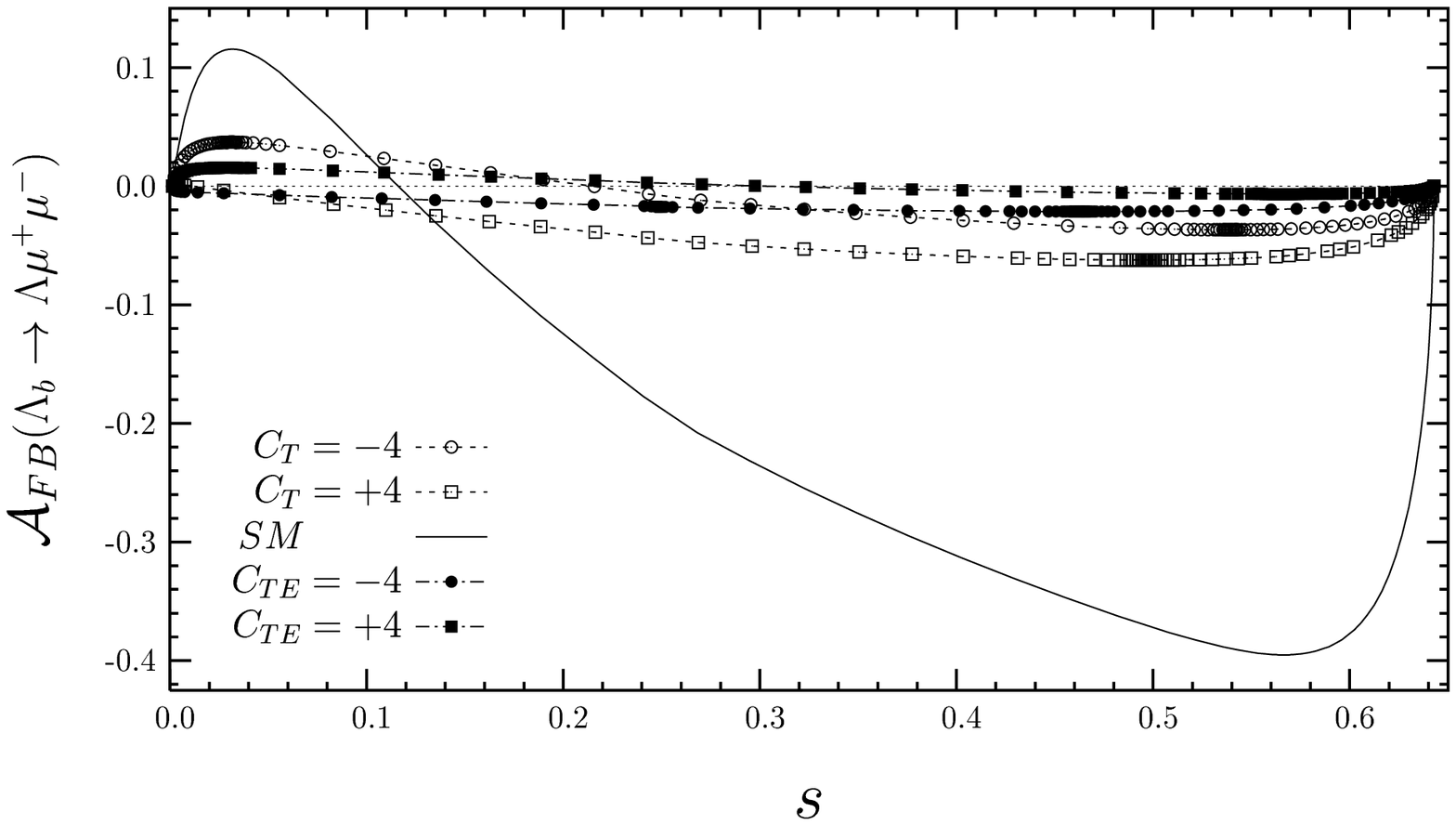}
\vskip 7.8 cm    
\caption{}
\end{figure}

\begin{figure}   
\vskip 1.5 cm
    \includegraphics{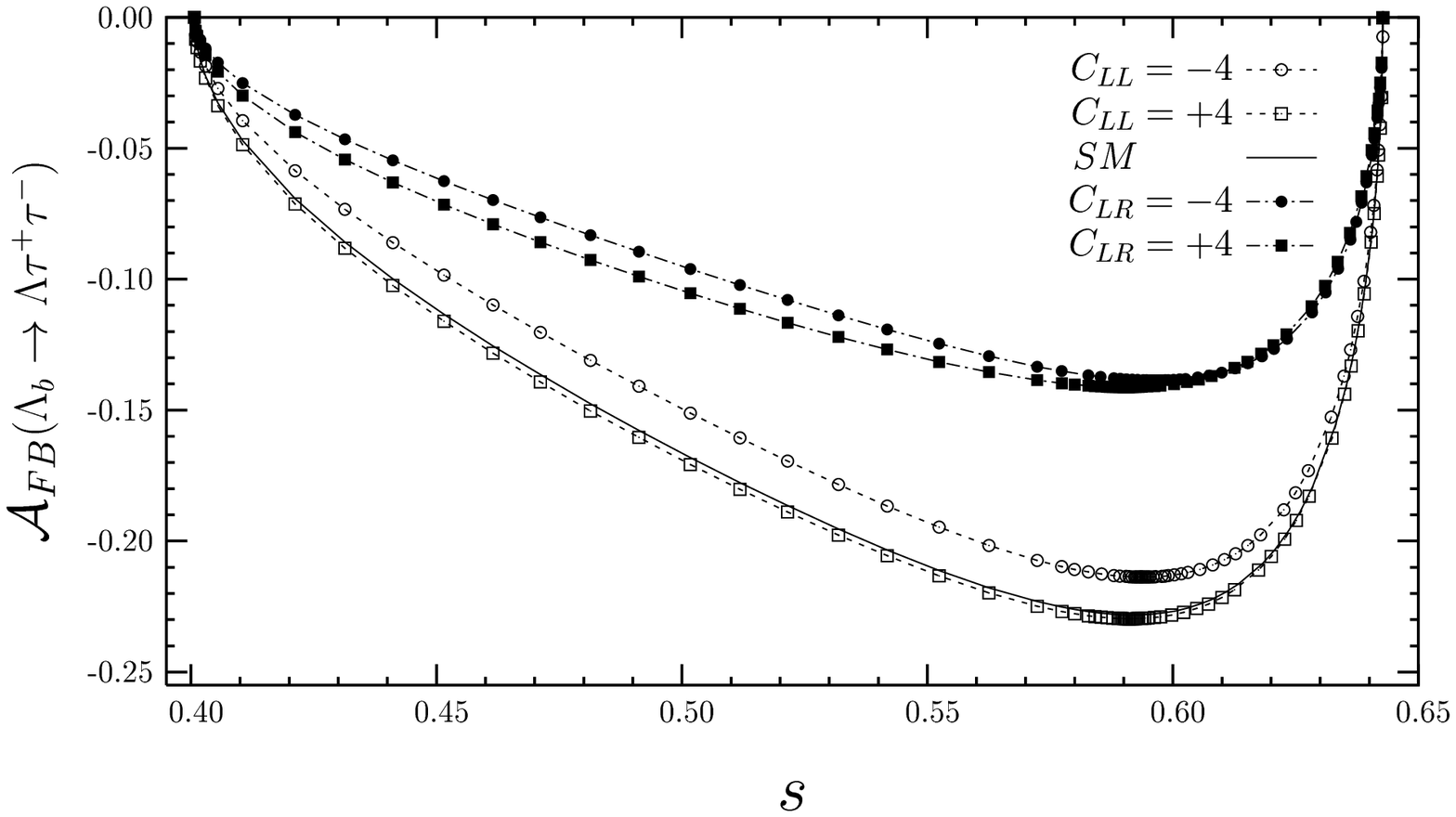}
\vskip 7.8cm
\caption{}
\end{figure}

\begin{figure}    
\vskip 2.5 cm
    \includegraphics{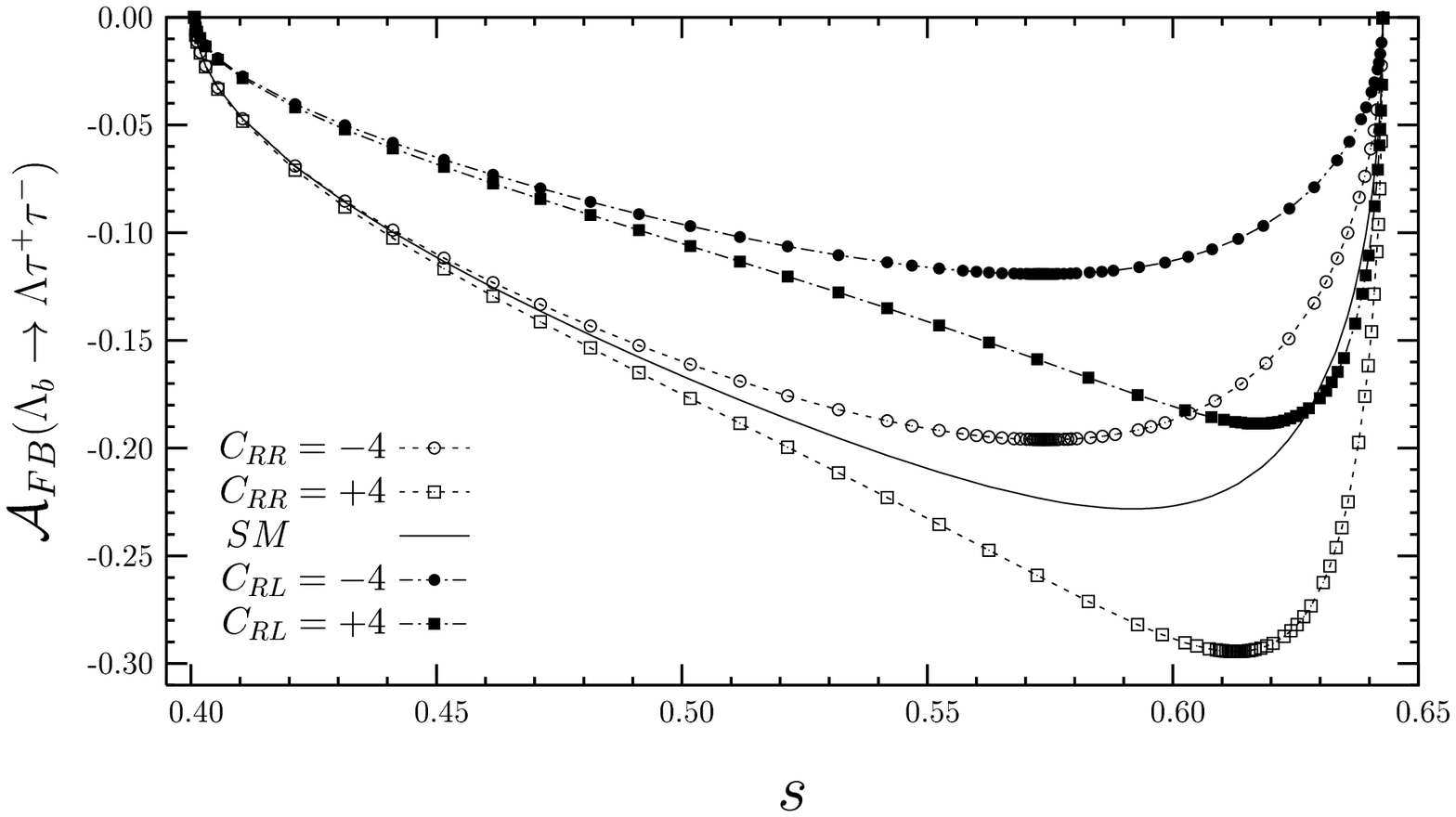}
\vskip 7.8 cm   
\caption{}
\end{figure}

\begin{figure}
\vskip 1.5 cm
    \includegraphics{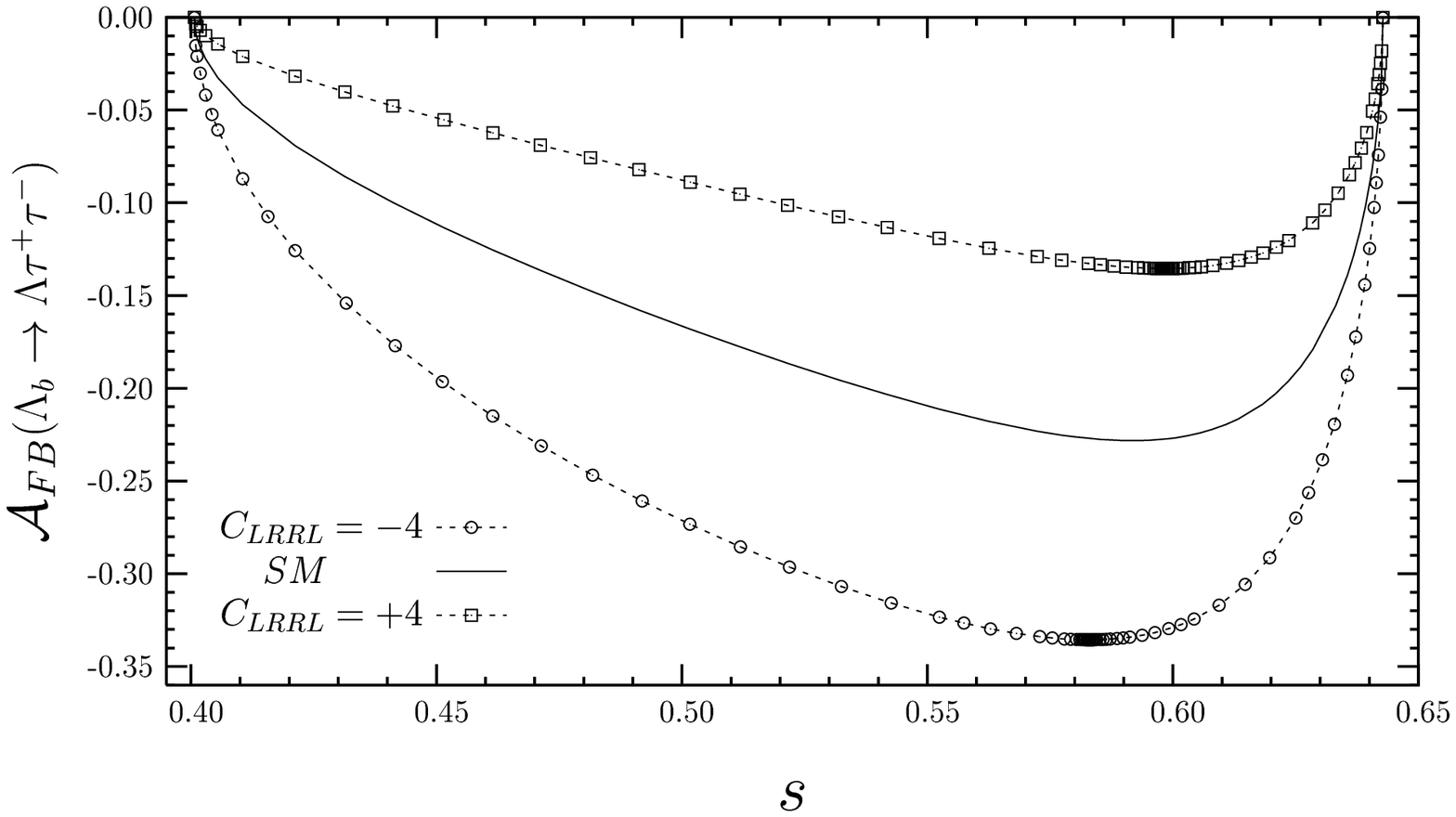}
\vskip 7.8cm
\caption{}
\end{figure}

\begin{figure}   
\vskip 2.5 cm
    \includegraphics{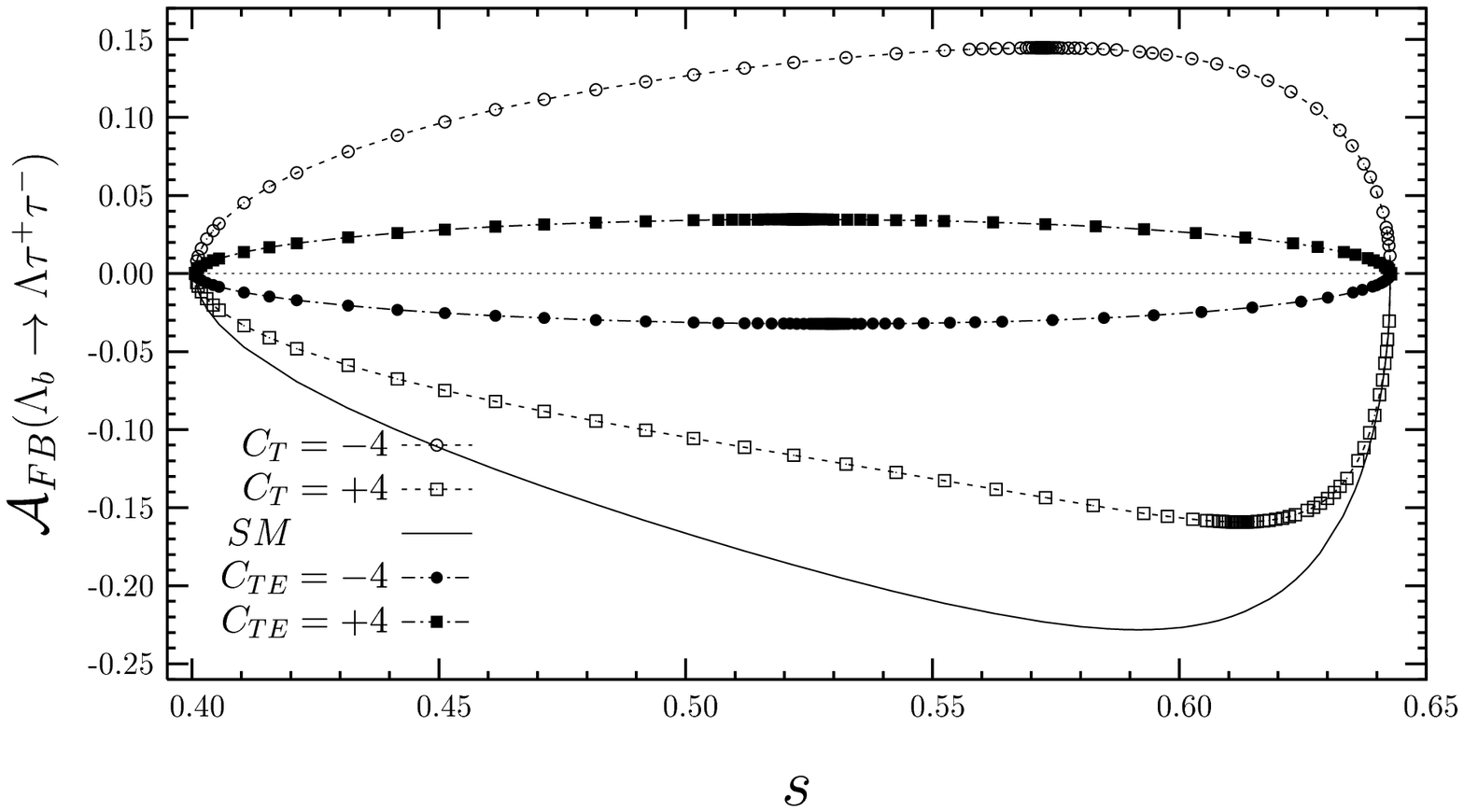}
\vskip 7.8 cm    
\caption{}
\end{figure}

\end{document}